\documentclass{aa}
\usepackage{graphicx,epsfig}
\usepackage{rotating}
\usepackage{supertabular}

\usepackage{natbib}
\bibpunct{(}{)}{;}{a}{}{,}
\usepackage{txfonts}
\def\vsini{$V\!\sin i$}
\def\teff{T$_ {\rm{eff}}$}

\def\logg{log~{\it g}}

\def\kps{km~s$^{\rm{-1}}$}

\def\lratio{R$_{AB}^\lambda$}
\def\cd{$d^{\rm -1}$}
\def\correction{}

\newcommand{\mcol}[3]{\multicolumn{#1}{#2}{#3} }

\newcommand{\struut}{\rule[-2ex]{0ex}{5.2ex}}
\newcommand{\struutup}{\rule{0ex}{3.2ex}}
\newcommand{\struutdown}{\rule[-2ex]{0ex}{2ex}}

\begin{document}
%

\title{Search for pulsation among suspected A-type binaries and\\
the new multiperiodic $\delta$ Scuti star HD~217860\thanks{This work is based on spectroscopic 
observations made at the Haute-Provence Observatory (OHP), the Observatoire du Pic du Midi (TBL)
and the Bulgarian National Astronomical Observatory (NAO, Rozhen). 
} }



\titlerunning{Search for pulsation among suspected A-type binaries}

   \author{Y. Fr\'emat\inst{1} \and P. Lampens\inst{1} \and P. Van Cauteren\inst{2}
   \and S. Kleidis\inst{3} \and K. Gazeas\inst{4} \and P. Niarchos\inst{4}
   \and\\ C. Neiner\inst{5} \and D. Dimitrov\inst{6} \and J. Cuypers\inst{1}
   \and J. Montalb\'an\inst{7} \and P. De Cat \inst{1} \and C.W. Robertson\inst{8}
   }

   \institute{Royal Observatory of Belgium,
             3 avenue circulaire, 1180 Brussel, Belgium\\
              \email{yves.fremat@oma.be}\\
              \email{patricia.lampens@oma.be}
   \and Beersel Hills Observatory, Beersel, Belgium
   \and Zagori Observatory, Epirus, Greece
   \and University of Athens, Department of Astrophysics, Astronomy and Mechanics
     Panepistimiopolis, 157 84, Zografos, Athens, Greece
   \and GEPI / UMR 8111 du CNRS, Observatoire de Paris-Meudon, 5
     place Jules Janssen, 92195 Meudon, France
   \and Institute of Astronomy, Bulgarian Academy of Sciences, 72 Tsarigradsko
     Shosse Blvd., 1784 Sofia, Bulgaria
   \and Universit\'e de Li\`ege, Institut d'Astrophysique et de G�ophysique,
     All�e du 6 Ao�t, 17 Sart Tilman, Li\`ege, Belgium
   \and SETEC Observatory, Goddard, Kansas, USA
   }



\abstract {In the H-R diagram, the intersection of the main
sequence and the classical Cepheid instability strip corresponds
to a domain where a rich variety of atmospheric phenomena are at 
play (including pulsation, radiative diffusion, convection). 
Main-sequence A-type stars are among the best candidates 
to study the complex interplay between these various phenomena.}
{We have explored a sample of suspected A-type binaries in a systematic
way, both spectroscopically and photometrically. The sample
consists of main-sequence A-type stars for which the few existing 
radial velocity measurements may show variability, but for which other 
essential information is lacking. Due to their location in the H-R diagram, 
indications of pulsation and/or chemical peculiarities among these suspected 
binary (or multiple) systems may be found.} 
{High-resolution spectroscopy obtained with the ELODIE and MUSICOS 
spectrographs was used in combination with a few nights of 
differential CCD photometry in order to search for pulsation(s). 
In order to search as well for chemical peculiarities or for
possible hidden component(s), we derived the atmospheric stellar 
parameters by fitting the observed spectra with LTE synthetic ones.}
{Of the 32 investigated targets,
{eight} are spectroscopic binaries, one {of which} is a 
close binary also showing eclipses, and three have been identified 
as $\delta$ Scuti pulsators with rapid line-profile variations.} 
{Among the latter stars, HD~217860 reveals interesting multiperiodic
photometric and spectroscopic variations, with up to {eight}
frequencies common to two large photometric data sets. 
{We suggest that at least one radial overtone mode is excited among 
the two most dominant frequencies, 
on the basis of the computation of the pulsation constants
as well as of the predicted frequencies and the expected behaviour of the 
amplitude ratio and the phase difference in two passbands using adequate 
theoretical modelling. We furthermore found evidence for a strong modulation 
of the amplitude(s) and/or the (radial) frequency content of this intriguing 
$\delta$ Scuti star.}}


   \maketitle
%

\section{Introduction}

In their catalogue of stellar radial velocities, \citet{1999A&AS..137..451G} noticed that
32\%~of the sample of B8-F2 type stars observed by the Hipparcos satellite have variable
velocities and assumed that this variability was due to multiplicity only. Though this
conclusion remains valid when performing statistics on the overall occurrence of multiplicity,
it is realistic to think that some of the studied stars are also non-radial pulsators showing
line profile variations (LPVs). In the H-R diagram, this situation may happen, for example,
at the intersection of the {\correction classical Cepheid instability strip} (CIS) and the main sequence
(where the mid-A to early-F type stars are).
In this region, a rich variety of phenomena are at play in the stellar atmospheres, some of
which are expected to produce long- and/or short-term variability. These phenomena consist
in the different pulsation mechanisms (active in the $\delta$ Scuti, SX Phe, $\gamma$ Dor and
roAp variable stars) and in various other processes involving magnetism, diffusion, rotation
and convection. The latter processes may boost or on the contrary inhibit the presence of
chemical peculiarities (occurring in Ap, Am, $\rho$ Pup and $\lambda$ Boo stars). The
competition between these processes and mechanisms thus leads to a large mix of stellar groups
of different atmospheric composition \citep{2004IAUS..224..499D}. These stellar groups also
behave in different ways with respect to pulsation and binarity, resulting in non-symmetric
spectral lines which can lead to misinterpretation of the radial velocity (RV) measurements.

Combining high-resolution spectroscopy and CCD photometry, the present work aims
at exploring the RV variability of several poorly known HIPPARCOS targets located at
the lower end of the instability strip. For one of the most promising targets (HD~217860) showing
multiperiodic variations, we performed a frequency analysis of the multi-site differential
photometric time series. The data and the reduction procedure are described in Sect.~\ref{sec:observations}, while the tools and the procedure we adopted to derive the stellar
parameters (effective temperature, surface gravity and projected rotational velocity) and their
errors are described in Sect.~\ref{sec:tools}. {Section~\ref{sec:results} and \ref{sec:individual}
provide respectively the global results and a discussion of several interesting targets. We present
the conclusions and future perspectives of this project in Sect~\ref{sec:conclusions}}.

\begin{table}
\caption{Target description. Nref is the number of references in
SIMBAD at present. The spectral type is from \citet{1999A&AS..137..451G}.
\label{tab:targets}}\center
\begin{tabular}{rrrrl}
\hline \hline\noalign{\smallskip}
HD      & HIP & Nref & V & Sp. Type \\
\hline\noalign{\smallskip}
849 & 1043 & 7 & 7.17 & A4 V\\
3066 & 2719 & 1 & 7.36 & A3 V\\
3743 & 3165 & 7 & 7.21 & A4 III\\
3777 & 3227 & 9 & 7.44 & A2 III\\
5066 & 4129 & 12 & 6.70 & A2 V\\
6813 & 5416 & 3 & 7.36 & A2 IV\\
7551 & 5886 & 3 & 6.71 & A3 V\\
11190 & 8581 & 1 & 7.88 & A2 III\\
12389 & 9501 & 8 & 7.98 & A4 V\\
12868 & 9851 & 3 & 7.25 & A4 II\\
13162 & 10045 & 1 & 7.92 & A2 IV\\
14155 & 10731 & 2 & 7.44 & A3 V\\
17217 & 13063 & 2 & 6.92 & A2 V\\
19257 & 14479 & 3 & 7.07 & A9 III\\
20194 & 15177 & 3 & 7.92 & A5 V\\
25021 & 18777 & 11 & 7.29 & A2 V\\
26212 & 19436 & 4 & 7.36 & A5 V\\
27464 & 20495 & 3 & 7.76 & A7 IV--III\\
30468 & 22352 & 6 & 7.03 & A2 IV\\
31489 & 22984 & 2 & 7.49 & A4 V\\
38731 & 27525 & 2 & 7.92 & A7 V\\
42173 & 29375 & 1 & 7.55 & A5 V\\
44372 & 30287 & 1 & 7.77 & A2 V\\
64934 & 38891 & 4 & 7.00 & A5 V \\
68725 & 40361 & 11 & 6.94 & F2 Ib$^{*}$\\
81995 & 46642 & 2 & 7.35 & A7 III \\
217860 & 113790 &2  & 7.30 & A8 III\\
221774 & 116321 & 1 & 7.38 & A4 IV\\
223425 & 117479 & 3 & 7.07 & A2 V\\
223672 & 117646 & 19 & 7.34 & A6 V\\
224624 & 118276 & 7 & 7.20 & A2 V\\
225125 & 300 & 1 & 7.45 & A7 IV\\
\hline\noalign{\smallskip}
\end{tabular}

\noindent
{\correction $^{*}$: rather a chemically peculiar main-sequence star (cf. Tab.~\ref{tab:stellarparameters})}
\end{table}

\section{Observations and data reduction}
\label{sec:observations}

\subsection{Target selection}

32 targets from the HIPPARCOS catalogue have been selected according to the following criteria :
1) brighter than magnitude 8; 2) spectral type ranging from A0 to F2; 3) showing
some indication of radial velocity variability \citep{1999A&AS..137..451G}; and 4)
{preferentially} with less than ten references in the bibliography recorded in the SIMBAD
database at the CDS. The selected targets are ordered by increasing HD number in Table~\ref{tab:targets},
with the HIP number (col. 2), number of references in SIMBAD (col. 3), V magnitude (col. 4)
and spectral type (col. 5) from \citet{1999A&AS..137..451G}.

\subsection{High-resolution spectroscopy}

The spectroscopic observations were carried out at the 1.93~m telescope of the OHP \citep{1996A&AS..119..373B}, equipped with the {\sc ELODIE} spectrograph (R$\sim$40000). High-resolution
spectra were collected during 4 nights in 2004 (December 3--7). Each target was observed
2 to 5 times in order to be able to detect rapid (periods of order of a few hours) or slow
(periods of the order of a few days) line profile variations (LPVs) and/or changes in radial velocity.
However, due to the weather conditions, some of our targets have only been observed 2--3
consecutive times, without the possibility to reobserve them at a later date. We adapted the time
exposures to ensure a S/N ratio per pixel usually varying from 70 to 100 (at 5000 \AA).
The journal of observations is given in Table~\ref{tab:journal} and contains:
the HD identifier (col.~1), Heliocentric Julian Day (HJD, col.~2), signal-to-noise
ratio (col.~3), exposure time (col.~4), as well as the instantaneous radial velocity (col.~5)
and the observatory's acronym (col.~6).

\begin{table}
\caption{Journal of spectroscopic observations at OHP, TBL and NAO. The
complete table is available in the electronic version of the
paper.\label{tab:journal}}\center
\begin{tabular}{rlrrr@{}r@{}rr}
\hline \hline\noalign{\smallskip}
HD      & HJD$-$2400000 & S/N & exp. & \multicolumn{3}{c}{RV} & Obs.\\
          &             &   & [s] & \multicolumn{3}{c}{[\kps]} &\\
\hline\noalign{\smallskip}
    849 & 53346.2338 & 137 &  1200 & 10.46&$\pm$&5.93 & OHP\\
        & 53346.2493 & 143 &  1200 & 5.17&$\pm$&4.10 & OHP\\
   3066 & 53343.3515 &  59 &   644 & $-$12.15&$\pm$&1.35 & OHP\\
        & 53343.3818 &  70 &  1200& $-$1.75&$\pm$&2.86 & OHP\\
        & 53344.3245 & 104 &  1200 & 2.68&$\pm$&4.17 & OHP\\
        & 53344.3399 & 104 &  1200 & $-$8.86&$\pm$&2.30 & OHP\\
\multicolumn{8}{l}{...}\\
\hline
\end{tabular}
\end{table}

\begin{figure}
\center
\includegraphics[width=8cm,clip=]{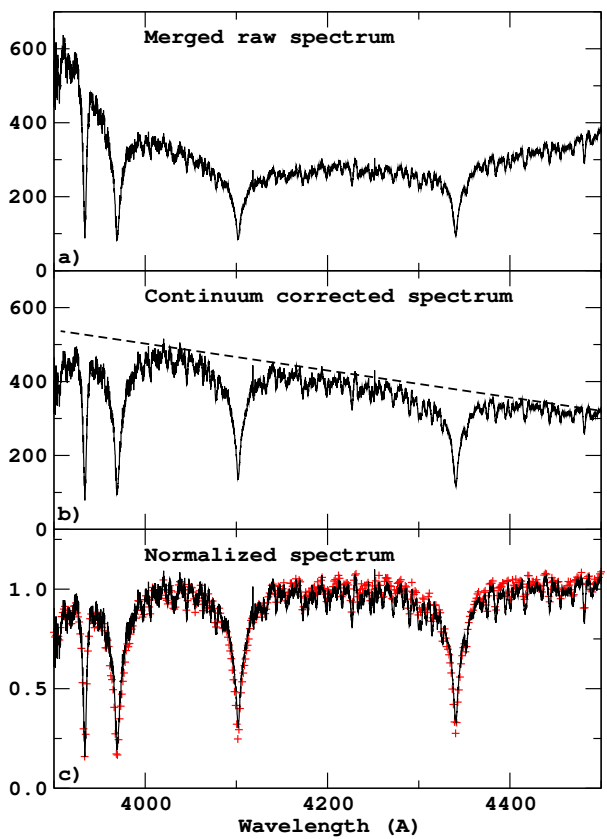}
\caption{Normalisation of the spectrum of HD~225125: a) Merged
spectrum with incomplete removal of the instrumental
signature/response; b) Corrected spectrum and continuum placement
(broken line); c) Normalized observed spectrum (black line)
compared to the synthetic one (red crosses).\label{fig:reduc}}
\end{figure}

The data have been automatically reduced order-by-order at the end of the night
using the {\sc INTERTACOS} pipeline. The first 50 echelle orders were merged using
the overlapping region and computing a ratio allowing to scale each order
as described in \citet{2002A&A...383..227E}. To ease normalisation of the spectra
and to correct for the incomplete removal of the instrumental response
(see Fig.~\ref{fig:reduc}a), we used the spectra of the stars in our sample with
known Str\"omgren and H$\beta$ colour indices. We adopted these indices to perform
a first estimate of the stellar parameters and to compute a reference synthetic spectrum.
An averaged function was obtained by dividing the observed data by this "reference
spectrum", which was further used to correct the merged raw spectra (see Fig.~\ref{fig:reduc}b).
Normalization was finally performed by fitting a line through the continuum between 4000 
and 4500 \AA~with the {\sc continuum} task of {\sc iraf} (see Fig.~\ref{fig:reduc}c).

To complete these data, some spectra were obtained with the {\sc Musicos} spectropolarimeter 
(R$\sim$35000) mounted at the Cassegrain focus of the 2-m telescope Bernard Lyot Telescope 
(TBL) at the {\it Observatoire du Pic du Midi} (France) \citep{1999A&AS..134..149D} in 
July 2005. The {\correction spectral domain covered} ranges from 4500 to 6600 \AA; these spectra 
were reduced with the ESpRIT software package developed by \citet{1997MNRAS.291..658D} 
and improved by \citet{2003A&A...411..565N}.

\begin{figure}
\center
\includegraphics[width=8cm,clip=]{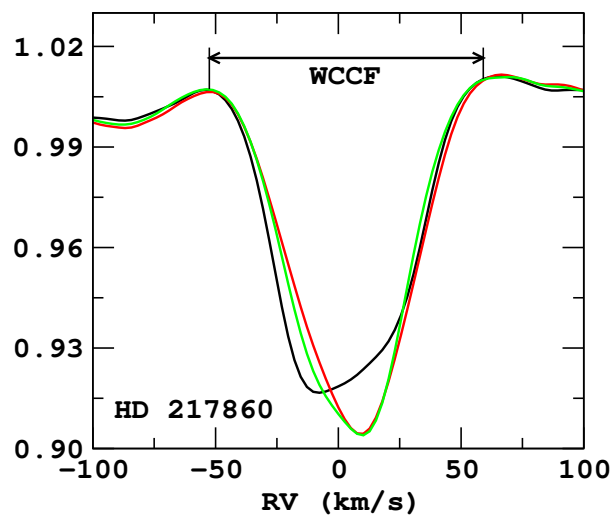}
\caption{The CCF of HD~217860 shown for 3 consecutive exposures.\label{fig:ccf}}
\end{figure}

{\sc INTERTACOS} and ESpRIT were also used to perform a cross-correlation of the observed
spectra with the appropriate mask after each exposure. However, since most of our targets 
have large \vsini~values, we recomputed the cross-correlation function (CCF)
in a spectral domain ranging from 5000 to 5700 \AA~in order to avoid the hydrogen lines. 
Synthetic spectra obtained with the stellar parameters derived in Sect.~\ref{sec:tools:parameters} 
and for \vsini~= 0~\kps~were adopted as templates. An example of a CCF can be found 
in Fig.~\ref{fig:ccf}.

{\correction In two cases (HD~11190 and HD~68725), additional spectra were obtained 
with the Coud\'e spectrograph (with a resolution of 0.19 \AA~/pixel) on the 2-m R-C
telescope of the NAO Rozhen in December 2006. The spectral domain covers three regions 
from about 4440 to 4640 \AA~, from 6300 to 6500 \AA~, and around the H$_{\alpha}$ line. 
These spectra were reduced with standard IRAF procedures. The corresponding radial 
velocities were measured with the cross-correlation technique using synthetic spectra.} 

\subsection{CCD photometry}

We additionally performed complementary and exploratory CCD
photometry for the most promising targets of the sample, i.e. for
those stars that showed interesting short-term variations of the
CCFs, in the period between December 2004 and January 2006. This
has not always been possible, however, as the targets are bright and
suitable comparison stars were not always available in the used
fields-of-view (FOVs). For this reason, various telescopes of
different size have been used, including also very small
instruments of only 13-cm aperture. Table~\ref{phot:instrum}
summarizes the technical specifications of these instruments. In
some cases we had no other choice than to use the only other
bright star in the field, even though the colour and/or magnitude
difference was not optimal. Depending on the target's magnitude, a
B or V filter according to Bessell's specifications
\citep{1995CCDA....2...20B} was employed.

\begin{table*}[ht]
\center \caption{Description of used instruments during the
exploratory program. FOV (i.e. col.~4) stands for "Field Of
View".\label{phot:instrum}}
\begin{tabular}{lccll}
\hline\hline\noalign{\smallskip}
Observatory  &    Telescope  &  Camera  & FOV &  Notes \\
\hline\noalign{\smallskip}
  BHO  &  13-cm refractor &  SBIG ST10XME & 44'{\rm x}29.5'& \\
       &  25-cm Newton    &  SBIG ST10XME & 34'{\rm x}23'  & f/6 \\
       &  40-cm Newton    &  SBIG ST10XME & 26'{\rm x}17.5'& f/5 \\
  HLO  &12.5-cm refractor &  SBIG ST10XME & 40.5'{\rm x}27.5'& guidescope of R-C tel. \\
  HLO  &1-m Cassegrain & {Enzian} & 24'{\rm x}12'& with focal reducer \\
\hline
\end{tabular}
\end{table*}

\begin{table*}
\caption{Journal of CCD photometric exploratory observations of
the variable candidates. Observations were made at the Beersel Hills
Observatory (BHO) and the Hoher List Observatory from the
Argelander Institute for Astronomy, Bonn (HLO). \label{phot:logbook}}
\begin{tabular}{@{}rrlccccl@{}} \hline
\hline\noalign{\smallskip} � HIP & HD & Date & Observat. & Telescope & Filter & Time span & Remarks \\
& �& �& �& Size�& �& (hours) & \\
\hline \noalign{\smallskip}
� �3165 � �& � 3743 & 12/13-Jan-05 & �BHO � & � 13-cm � & � V � & �2 � & variable (with a periodicity of days ?)\\
� � � � � �& � � � �& 13/14-Jan-05 & �BHO � & � 13-cm � & � V � &
�2 � & shows a difference of 0.078~mag in mean light level\\
\noalign{\smallskip}
� �9501 � �& �12389 & 19/20-Dec-04 & �BHO � & �25-cm �& � V � & �
4.8 � & previously known $\delta$ Scuti variable\\
\noalign{\smallskip}
� 15177 � �& �20194 & 10/11-Dec-05 & �HLO � & �12.5-cm �& � V � & �6.3 �& constant ($\sigma$=6 mmag)\\
� � � � � �& � � � �& 14/15-Jan-06 & �BHO � & �25-cm �& � V � & �6 & constant ($\sigma$=5 mmag)\\
� 40361 � �& �68725 & 12/13-Jan-05 & �BHO � & � 40-cm � & � B � & � 7 � & new $\delta$ Scuti variable\\
� � � � � �& � � � �& 10/11-Dec-05 & �HLO � & �12.5-cm �& � B � &
� 5 � & \\ \noalign{\smallskip}
� 46642 � �& �81995 & �1/2-Apr-05 �& �BHO � & � 40-cm � & � B � & � 6 � & partial eclipse; $\Delta$m(variable$-$check) $\approx$ 0.06 mag\\
� � � � � �& � � � �& 10/11-Apr-05 & �BHO � & � 40-cm � & � B � & � 5.3 � & partial eclipse \\
� � � � � �& � � � �& 11/12-Apr-05 & �BHO � & � 40-cm � & � B � &
� 4.8 �& constant ($\sigma$=4 mmag)\\\noalign{\smallskip} �
 113790 � & 217860 & �3/19/20-Dec-04 & �BHO � & � 40-cm � &
� B � & 18.4 �& new $\delta$ Scuti variable\\
\noalign{\smallskip}
� 116321 � & 221774 & �6/7-Feb-05 �& �HLO � & �1-m � & � B � & � 4.1 �& constant ($\sigma$=4 mmag)\\
� � � � � �& � � � �&    14/15-Jan-06 & �BHO � & � 40-cm � & � B �
& � 7 �& constant ($\sigma$=4 mmag)\\ \noalign{\smallskip} \hline
\end{tabular}
\end{table*}

The most promising candidates for short-term variability in the CCFs were
submitted to two photometric runs of about half a night each
in order to verify the presence of short-periodic light variations.
Table~\ref{phot:logbook} summarizes the journal of the observations
for 7 of the 32 selected HIPPARCOS targets. In one additional case (HD~30468), 
{there was no suitable} comparison stars in the field.

\subsection{Multi-site CCD photometry for HD~217860}

HD~217860 was recognised by HIPPARCOS as a variable star but with an
unsolved variability pattern \nocite{1997yCat.1239....0E} (ESA
1997). {\correction Much to our surprise, a brief frequency-analysis of 
113 selected measurements from the Hipparcos Epoch Photometry allowed
us to detect a most dominant frequency at 19.7474 \cd with a significance
ratio of about 10 (Hp-amplitude of 18 mmag), accompanied by a possible 
second frequency at 14.2681 \cd with a significance ratio between 5 and 6 
(Hp-amplitude of 10 mmag)}. The variations in the light curves and the spectra 
are rapid \citep[see Fig.~\ref{light:curves} and also][]{2006CoAst.148...77F} 
and show the presence of multiple periods, confirming its status of $\delta$ 
Scuti variable star. {\correction The light curves are sometimes very peculiar.} 
We therefore collected CCD photometric (differential) data from
various observatories equipped with small instruments in Europe,
in the period between December 3, 2004 and December 25, 2005. The
observations were carried out using a 0.4-m Newton equipped with
a SBIG ST-10 XME CCD at Beersel Hills Observatory (BHO,
Belgium), a 0.4-m reflector with a focal reducer and SBIG ST-8
CCD at the University of Athens Observatory (UAO, Athens,
Greece), and a 20-cm telescope with a ST-7 XMEI CCD at Athens
(ZO, Greece). We performed the observations using the B
and V filters according to Bessell's specifications (Bessell 1995).
Table~\ref{tab:log} shows the details of the observational
campaigns. Due to {\correction a technical problem, observations obtained at} a fourth 
site could however not be used. We employed the software Mira AP\footnote{The Mira AP
software is produced by Mirametrics Inc.} (vers. 6) to reduce the
images following standard procedures (offset and dark current correction,
flat-field calibration) and to compute the differential magnitudes
using the technique of aperture photometry at BHO. At ZO and UAO, we
used the package AIP4WIN (respectively the versions 1.4.21 and 1.4.25)
\citep{2005haip.book.....B}.

All differential magnitudes have been computed using HIP~113918 (GSC~3997:1091) as the
principal comparison star and GSC~3997:1078 as the check star. At BHO, additional comparison
stars have been considered. In this case we identified GSC 3997:775 as NSV~14402 in the
same field and also measured it. Since both HIP~113918 (the comparison star 'C1') and HIP~113790
(the variable star 'V') are much brighter than GSC~3997:1078 (the check star 'K'), the standard
deviations of the mean differential magnitude in the sense (K - C1) are used as estimates
of the highest noise level expected to be found in the residual data during the subsequent
frequency analyses. We subtracted the overall averaged values for every measured star at each
observatory. We initially also corrected for nightly shifts using the mean values of the
differential magnitudes in the sense (K - C1) per night and per observatory.
We next computed the relative weights of each time series by determining the night-to-night
standard deviations of the (K - C1) differences, and we adopted a weight
inversely proportional to their variance. The largest relative weight (set equal to 1) has been
taken from the highest-quality time series. The preliminary analyses however showed that the
corrections based on the (K - C1)'s nightly averages did not compensate
well enough for the nightly shifts of the (V - C1) data in order to be able
to remove most of the power in the low frequency regime. Therefore, we went one step further and
corrected for nightly shifts using the mean values of the differential magnitudes in the sense
(V - C1) per night and per observatory. In this way, we removed any signal
that might indicate a real long-term periodical trend in the datasets. However, since our main
interest is the study of the pulsations, we will only focus on the high-frequency regime from hereon.

\begin{table*}[t]
\begin{center}
\caption{Log of the CCD differential multi-site photometric campaign conducted for HD~217860.\label{tab:log}}
\begin{tabular}{cccrccc}
\hline
\hline
Dates (yr 2005) & Site & Observer(s) & N$_{data}$ &  C1 & K & NSV\struut \\
\hline
\mcol{1}{c}{\it Filter B} &&&&&&\\
Nov.  4 -- Dec.  25 & UAO & KG & 1274 & y & y & n \\
Dec.  3, 2004 -- Nov.  22 & BHO & PVC, PL &  2970 & y & y & y \\
Jun. 13 -- Dec.  25 & ZO  & SK  & 5667 & y & y & n \\
Dec.  3, 2004 -- Dec.  25 & All & All & 9911 & y & y & n  \\
\hline
\mcol{1}{c}{\it Filter V} &&&&&&\\
Nov.  4 -- Dec.  25 & UAO & KG & 1130 & y & y & n  \\
Jul. 14 -- Nov.  22 & BHO & PVC, PL &  2751 & y & y & y \\
Jul.  5 -- Dec.  25 & ZO  & SK  & 1599 & y & y & n  \\
Jul.  5 -- Dec.  25 & All & All & 5384 & y & y & n  \\
\hline
\end{tabular}
\end{center}
\end{table*}

\section{Procedure and tools}
\label{sec:tools}

{\correction
We derived the fundamental stellar parameters (\teff, \logg~and \vsini) by fitting our
high-resolution spectra with synthetic ones (Sect. \ref{sec:tools:parameters}). For the
double-lined spectroscopic binaries (SB2s) of our sample, the procedure was adapted to 
account for the existence of a companion and an additional parameter (R$_{AB}^\lambda$), 
representing the monochromatic luminosity ratio between the components A and B, was defined 
(Sect. \ref{sec:tools:parameterssb2}).
The error bars for R$_{AB}^\lambda$, \vsini~and \teff~represent the standard errors 
resulting from the different fitted zones, while those for \logg~are derived from the 
uncertainties on the parallax (and on R$_{AB}^\lambda$ for SB2s), on the one hand, 
and on the effective temperature, on the other hand.
}

\subsection{Stellar parameter determination: the general case}
\label{sec:tools:parameters}

The radial velocity (RV), projected rotational velocity (\vsini), effective temperature (\teff),
and surface gravity have been derived in four consecutive steps. First, the CCFs were used to estimate
the instantaneous and averaged RV. Then, a comparison between observed and synthetic spectra allowed
us to derive the projected rotational velocity and the effective temperature (Sect.~\ref{sec:tools:models}).
This has been performed by means of the {\sc girfit} computer code \citep{2005astro.ph..9336F}, which
allows to conduct a least squares fitting based on the {\sc minuit} minimization package.
In this study, we mainly focused on several zones in the spectral domain ranging from 3900 to
4500 \AA~(Table~\ref{tab:zones}), in which {independent} fits were performed.
The determination of \vsini~is based on metallic line fitting with RV fixed and considering
\logg, \teff, and \vsini~as free parameters. It is worth to remark that, at this stage,
the procedure provides \vsini~only, while the values obtained for \teff~and \logg~are not yet
meaningful. Since most of the targets are cooler than 8500~K, the determination of \teff~was
performed using the hydrogen lines as well as the \ion{Ca}{ii} K-line. This spectra-fitting
{procedure} was completed adopting the previously obtained values of \vsini~and RV, while
\logg~was kept equal to 4.0 in the first iteration. The surface gravity has finally been derived by
combining the HIPPARCOS parallax, V magnitude and \teff~value in order to estimate the
stars' luminosity \citep{2003A&A...398.1121E}. We then obtained the mass and radius
from theoretical stellar evolutionary tracks computed for Z=0.02 \citep{1992A&AS...96..269S}.
A second iteration has been performed to test the sensitivity of the \teff~determination to a change in
\logg. 


\begin{table}
\caption{Fitting zones used to derive the projected rotation
velocity and effective temperature.\label{tab:zones}} \center
\begin{tabular}{cl}
\hline \hline\noalign{\smallskip}
\multicolumn{2}{c}{1. \vsini~determination}\\
zone & line-type\\
\hline\noalign{\smallskip}
4200 -- 4230 & Metals\\
4230 -- 4260 & Metals\\
4260 -- 4290 & Metals\\
4445 -- 4460 & Metals\\
4460 -- 4475 & Metals\\
4475 -- 4487 & \ion{Mg}{ii}\\
4485 -- 4500 & Metals\\
\hline\noalign{\smallskip}
\multicolumn{2}{c}{2. \teff~determination}\\
zone & line--type\\
\hline\noalign{\smallskip}
3915 -- 3950 & \ion{Ca}{ii} K--line\\
3950 -- 4000 & H$\epsilon$\\
4010 -- 4200 & H$\delta$\\
4200 -- 4270 & Metals\\
4270 -- 4400 & H$\gamma$\\
4450 -- 4500 & Metals\\
\hline\noalign{\smallskip}
\end{tabular}
\end{table}

\subsection{Stellar parameter determination in the case of SB2s}
\label{sec:tools:parameterssb2}

Stellar parameter determination of three SB2s of the sample (HD~6813, HD~11190, HD~221774)
was carried out by enabling the {\sc girfit} programme
to combine two wavelength shifted normalized synthetic spectra.
In this version of the code, the input parameters are \teff, \logg, \vsini, and RV for each
component. We further added a parameter representing
the monochromatic luminosity ratio, R$_{AB}^\lambda$=L$_{A, \lambda}$/L$_{B, \lambda}$,
between the components A and B of the system. RV and \vsini~values of each component
have been derived in the same way as for single stars. In a first step, the effective temperature
and the luminosity ratio of the stars have been fitted simultaneously, keeping the values
of \logg, RV, and \vsini~fixed. The value of R$_{AB}^\lambda$ is then used to estimate the
components' V magnitude which, together with the parallax and the effective temperature,
enables to derive the surface gravity (see Sect.~\ref{sec:tools:parameters}).
Where needed, several iterations were performed to get a coherent set of parameters.

\subsection{Model atmospheres and flux grid}
\label{sec:tools:models}

In order to determine the stellar parameters, the {\sc girfit} code computes the synthetic 
spectra from a \teff / \logg-interpolation in a grid of fluxes created with the {\sc 
synspec} programme \citep[][ see references therein]{1995ApJ...439..875H}. To account
for additional opacities due to Rayleigh scattering and H$^{-}$ ions, we enabled the 
IRSCT and IOPHMI opacity flags of {\sc synspec}. All the calculations were performed with 
{\sc atlas 9} using LTE atmosphere models computed by \citet{2003IAUS..210P.A20C}. 
The microturbulent velocity was supposed to be 2 \kps~and a solar-type chemical 
composition was considered.

{The programme {\sc spectrum} \citep{spectrum}
as well as the previously mentioned grid of LTE atmosphere models (for a solar-type chemical 
composition) 
were used to compute the synthetic spectra for obtaining the additional NAO radial velocities.}


\begin{table*}
\center \caption{Stellar parameters of the targets. Remarks: "VAR"
are the short-term variables; "ell." means ellipsoidal variations
are detected in the CCD photometry. The spectra of the 3 framed SB2 targets were analysed accounting
for the contribution of two components. {\correction Numbers between brackets represent the number of measurements 
used to compute the average radial velocity (col.~6).} \label{tab:stellarparameters}}
\center
\begin{tabular}{r@{}lr@{}lr@{}r@{}rr@{}r@{}rr@{}r@{}rr@{}r@{}rl}
\hline \hline\noalign{\smallskip}
HD     & & HIP & & \multicolumn{3}{c}{\teff} & \multicolumn{3}{c}{\logg} & \multicolumn{3}{c}{\vsini} & \multicolumn{3}{c}{RV} & Remarks \\
        & &   & & \multicolumn{3}{c}{[K]} & \multicolumn{3}{c}{ } & \multicolumn{3}{c}{[\kps]} & \multicolumn{3}{c}{[\kps]} & \\
\hline\noalign{\smallskip}
   849  && 1043   && 7714 &$\pm$& 285 & 3.99     &$\pm$& 0.07      & 169 &$\pm$& 10  & 7.33    &$\pm$&7.21 (2) & \\
  3066  && 2719 && 8149 &$\pm$& 450 & 3.97  &$\pm$&  0.09     & 255 &$\pm$& 20  & $-$7.26 &$\pm$& 5.72 (4)  & \\
  3743  && 3165 && 7895 &$\pm$& 3.81  & 3.81 &$\pm$&  0.15     & 97 &$\pm$& 3 & $-$8.16 &$\pm$& 1.88 (3)  & VAR, ell. ?, SB2 \\
  3777  &A& 3227      &A & 8242 &$\pm$& 180 & 4.04 &$\pm$&  0.10     & 12 &$\pm$& 1 &          34.06&$\pm$& 0.41 (2)        & SB2, Am \\
    5066  && 4129     &  & 9336 &$\pm$& 246 & 3.58 &$\pm$& 0.09 & 127 &$\pm$& 5 & $-$10.45&$\pm$&  5.24 (2) &  \\
  \hline
  \multicolumn{1}{|r@{}}{6813}  && 5416 &&  \multicolumn{12}{l}{$\gamma$= 13 $\pm$ 10 \kps; \lratio~= 4.36$\pm$ 1.56}& \multicolumn{1}{l|}{SB2}\\
  \multicolumn{1}{|r@{}}{6813}  &A& 5416      &A& 8016 &$\pm$& 350    &          & &       & 69 &$\pm$& 10 &         & &         & \multicolumn{1}{r|}{}\\
  \multicolumn{1}{|r@{}}{6813}  &B& 5416      &B& 8596 &$\pm$& 350    &          & &       & 10 &$\pm$& 2 &         & &         & \multicolumn{1}{r|}{}\\
  \hline
  7751  && 5886     & & 8240 &$\pm$& 186 & 3.90     &$\pm$&   0.10    & 157 &$\pm$& 6   & 11.21   &$\pm$&6.80 (2) &    \\
 \hline
 \multicolumn{1}{|r@{}}{11190}  && 8581 &&  \multicolumn{12}{l}{$\gamma$= 2.30$\pm$0.47 \kps; \lratio~= 2.02$\pm$ 0.02} & \multicolumn{1}{l|}{SB2}\\
 \multicolumn{1}{|r@{}}{11190}&{A}& {8581} &{A}& {7519} &$\pm$& {150} & {3.80} &$\pm$& {0.19} & {16} &$\pm$& {2} &  & &   & \multicolumn{1}{l|}{Am} \\
 \multicolumn{1}{|r@{}}{11190}  &{B}& {8581} &{B}& {7416} &$\pm$& {230} & {4.02} &$\pm$& {0.19} & {11} &$\pm$& {2} &   & &   & \multicolumn{1}{l|}{Am}\\
 \hline
 12389  && 9501   && 8155 &$\pm$& 250 & 3.64     &$\pm$& 0.20  & 82  &$\pm$& 5   & $-$37.84&$\pm$&2.00 (2) &  known $\delta$ Sct, VAR\\
 12868 & & 9851   && 8236 &$\pm$& 164 & 3.26     &$\pm$& 0.21 & 7   &$\pm$&0.6  & $-$3.89 &$\pm$&0.16 (2) & SB1? \\
 13162  && 10045  && 9104 &$\pm$& 170 & 3.85     &$\pm$&  0.18     &  27 &$\pm$& 2   & $-$5.20 &$\pm$&0.37 (2) &  \\
 14155  && 10731 & & 8300 &$\pm$& 350 & 4.19  &$\pm$&  0.09     & 234 &$\pm$& 18 & $-$11.09 &$\pm$& 10.32 (2)  & VAR \\
 17217  && 13063  && 8610 &$\pm$& 200 & 4.13     &$\pm$&   0.07    & 216 &$\pm$& 6   & $-$9.53 &$\pm$&7.60 (2) &  \\
 19257  && 14479 & & 7367 &$\pm$&195  & 4.23     &$\pm$&  0.06     & 135 &$\pm$& 6   & 0.26    &$\pm$&4.88 (2) & VAR \\
 20194  && 15177  && 7702 &$\pm$&200  & 4.03     &$\pm$&  0.12     & 238 &$\pm$&18   & $-$5.98 &$\pm$&6.18 (2) & \\
 25021  && 18777  && 8056 &$\pm$& 50  & 3.99     &$\pm$&  0.10     &85   &$\pm$&4    & $-$23.28&$\pm$&2.36 (2) & Am \\
 26212  && 19436  && 7873 &$\pm$& 290 & 4.09     &$\pm$&  0.10     & 221 &$\pm$& 14  & 9.67    &$\pm$&11.18 (4)& \\
 27464  && 20495  && 7470 &$\pm$& 175 & 3.18     &$\pm$&  0.22     & 105 &$\pm$& 5   & $-$1.30 &$\pm$&3.25 (2) & \\
 30468  && 22352 && 9068 &$\pm$& 213 & 3.98     &$\pm$&  0.10     & 129 &$\pm$& 7   & $-$12.29&$\pm$&7.25 (3) &  \\
 31489  && 22984 && 7751 &$\pm$& 350 & 3.91     &$\pm$&   0.11    & 236 &$\pm$& 17  & 15.17   &$\pm$&4.41 (2) & \\
 38771  && 27525 && 7753 &$\pm$& 220 & 3.28     &$\pm$&  0.25     & 79  &$\pm$& 5   & 7.47    &$\pm$&2.71 (2) &  \\
 42173  && 29375 && 7590 &$\pm$& 148 & 4.09     &$\pm$&   0.09    & 194 &$\pm$& 18  & $-$2.46 &$\pm$&8.25 (3) & \\
 44372  && 30287 && 8386 &$\pm$& 400 & 3.60      &$\pm$&  0.25     & 107 &$\pm$& 16  & 12.42   &$\pm$&2.00 (1) & \\
 64934  && 38891 && 7849 &$\pm$& 371 & 4.00     &$\pm$&  0.10     & 227 &$\pm$& 26  & $-$14.67&$\pm$&5.36 (2) &   \\
 68725  &&  40361 && 7000 &$\pm$& 100 & 3.60     &$\pm$&  0.11     & 41  &$\pm$& 4   & $-$5.80  &$\pm$&6.91 (10)& SB2 ? or/and CP/Am, new $\delta$ Sct, VAR\\
 81995  && 46642 && 7868 &$\pm$& 144 & 3.97     &$\pm$&   0.10    & 62  &$\pm$& 2   & 24.04   &$\pm$&2.18 (4) & VAR, SB1, eclipsing \\
217860  && 113790 && 7286 &$\pm$& 100 & 3.87     &$\pm$&  0.05     & 32  &$\pm$& 2   & 3.58    &$\pm$&2.45 (3) & new $\delta$ Sct, VAR \\
\hline
 \multicolumn{1}{|r@{}}{221774} && 116321 &&  \multicolumn{12}{l}{$\gamma$= 13 $\pm$ 5 \kps; \lratio~= 3.2$\pm$ 1.2} & \multicolumn{1}{l|}{SB2}\\
 \multicolumn{1}{|r@{}}{221774}  &A& 116321 &A& 8056 &$\pm$& 200 &          &   &       & 41 &$\pm$& 1 &         & &         &  \multicolumn{1}{r|}{}\\
 \multicolumn{1}{|r@{}}{221774}  &B& 116321 &B& 7132 &$\pm$& 400 &          &   &       & 73 &$\pm$& 11 &         & &         &  \multicolumn{1}{r|}{}\\
 \hline
223425  && 117479 && 8712 &$\pm$& 268 & 4.23 &$\pm$&  0.08     & 192 &$\pm$& 17  & 1.83   &$\pm$& 2.82 (2)  & \\
223672  && 117646 && 8015 &$\pm$& 237 & 3.30 &$\pm$&  0.34     & 164 &$\pm$& 9   & 8.05  &$\pm$& 4.60 (4)   & \\
224624  && 118276 && 8439 &$\pm$& 256 & 4.21 &$\pm$&  0.07     & 189 &$\pm$& 14 & $-$9.34 &$\pm$& 1.94 (3) & \\
225125  && 300 && 7631 &$\pm$& 235 & 3.93 &$\pm$&  0.10     & 118 &$\pm$& 4 & $-$4.70 &$\pm$& 2.02 (2) & \\
\hline
\end{tabular}
\end{table*}

\begin{figure}
\center
\includegraphics[width=8cm,clip=]{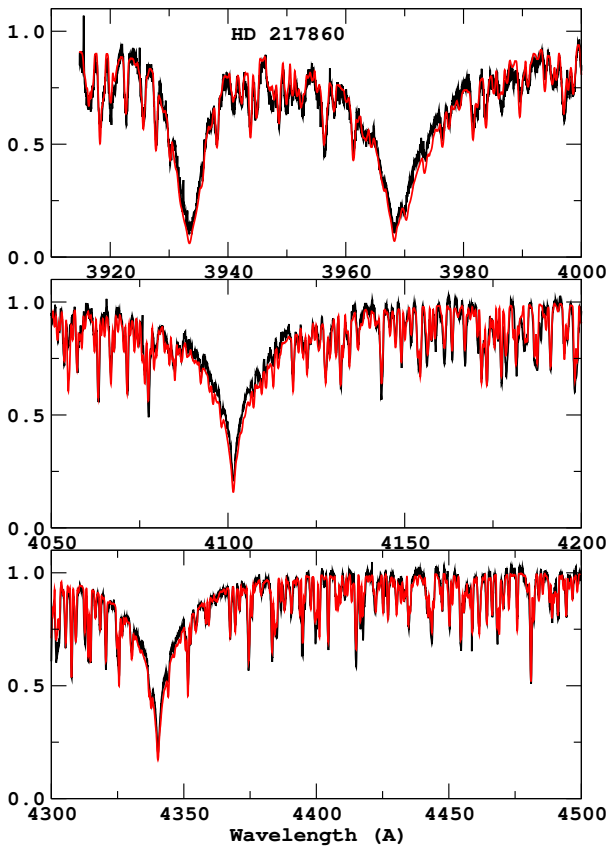}
\caption{HD 217860 -- Comparison between observed (dotted black
line) and synthetic spectra (full red line).\label{fig:o1sp}}
\end{figure}

\section{Results}

\label{sec:results}

\subsection{Spectroscopy}

\begin{figure}
\center
\includegraphics[width=8cm,clip=]{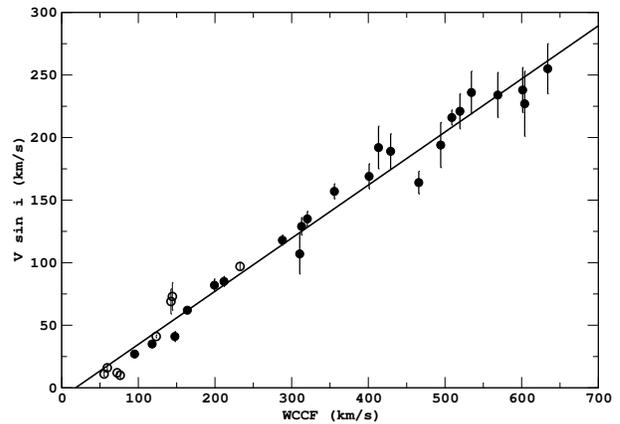}
\caption{Relation between the full base width of the
cross-correlation function (WCCF) and the \vsini~parameter. The
linear least squares relation between the measurements made on
single-lined objects (filled circles) is represented by the full
line. \vsini~values obtained for components member of multiple
systems are also plotted (unfilled circles).\label{fig:loi}}
\end{figure}

The procedure described in Sect.~\ref{sec:tools} was applied to the spectra
of all the targets of our sample. As an example of the agreement we obtained between
observed and synthetic spectra, we show those of HD~217860 in Fig.~\ref{fig:o1sp}.
The resulting stellar parameters are listed for each target in Table~\ref{tab:stellarparameters}:
the HD (col.~1) and HIP (col.~2) identification numbers, the effective temperature
(col.~3), the surface gravity (col.~4), the projected rotation velocity (col.~5), and
the radial velocity (col.~6). Remarks concerning multiplicity, pulsation and/or chemical
composition are reported in col.~7. In Fig.~\ref{fig:loi}, the projected rotation
velocities of single stars and of single-lined spectroscopic binaries (SB1s) are
plotted relative to the full {width of the estimated baseline of} the CCF (WCCF, see Fig.~\ref{fig:ccf}).
The linear relation that exists between WCCF and \vsini~can be used to verify the projected
rotation velocity determination of stars belonging to spectroscopically resolved
binaries (SB2s).


\subsection{CCD photometry}

\begin{figure}
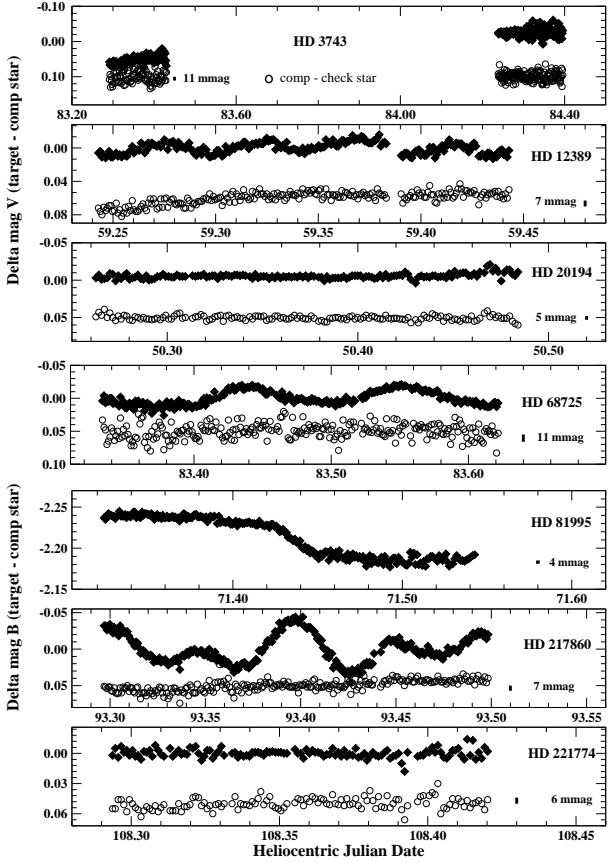

\center
\includegraphics[width=8cm,clip=]{3figH15177new.eps}
\includegraphics[width=8cm,clip=]{figH40361new.eps}
\includegraphics[width=8cm,clip=]{3figH116321new.eps}
\caption{\correction CCD light curves (filter V or B) obtained for 
seven targets having an obviously variable cross-correlation 
function. Magnitude differences (i.e target minus comparison 
star and comparison minus check star) are reported as a function 
of the Julian date. Arbitrary shifts were applied to present the 
light curves of the target and comparison data in a single panel.
In the case of HD 81995, there was no suitable check star available
in the field. Small thick lines show the corresponding standard deviations 
of the reference data obtained under the best conditions.\label{light:curves}}
\end{figure}

Fig.~\ref{light:curves} illustrates the light curves obtained in the B or V filter
for seven targets: HD~3743 (a new binary), HD~12389 (a known $\delta$ Scuti star),
{HD~20194 (a photometrically constant star)}, HD~68725 (a newly detected $\delta$
Scuti star), HD~81995 (a new eclipsing binary), HD~217860 (a new multiperiodic
$\delta$ Scuti star) and HD~221774 (a photometrically constant SB2). The discovery of
two new $\delta$ Scuti variable stars has already been reported elsewhere \citep{2005CoAst.146....6F}.
We generally used an arbitrary shift to plot the light curves of both the {\it target minus
comparison star} and the {\it comparison minus check star} in one and the same panel (after
having removed the mean difference in magnitude for each data set).

\subsection{Frequency-analysis for HD~217860}\label{sec:freq}

\subsubsection{Results}
\label{sec:freqres}

We performed the frequency analyses with the software package
{\sc Period04} which is based on the classical Fourier analysis
\citep{2005CoAst.146...53L}.
As a first step, we carried out a frequency search on the B- and the
V-time series for each observatory individually. These preliminary
{\correction runs indicated two common frequencies in each data set}. We next
{\correction merged} the data from all three observatories into larger sets of
weighted  measurements, one for each filter (sets 'B All' and 'V All').
First we computed the respective spectral  windows {\correction showing 
the alias features caused by the gap of almost one year ($\approx$ 0.003 \cd) 
and the 1 $d^{-1}$ spacing}. The subsequent frequency searches were performed 
on the weighted data (using the option  'Point Weight') in the interval from
0 to 35 \cd with a frequency step always smaller than  1.5{\rm x}10$^{-4}$ in 
B and 2.5{\rm x}10$^{-4}$ in V.
The total time span is 387 days, which corresponds to 0.0025 \cd~in frequency, 
for set 'B All' and 158 days, which corresponds to 0.0063 \cd, for set 'V All'. 
After each computation, the most dominant frequency was prewhitened  from the 
original (respectively the residual) data in successive steps of the frequency 
searches.  We stopped the analyses when a significance of 4 above the binned noise 
level (for the amplitude of an adopted frequency) in the periodogram of the residuals 
was reached {\correction or when the incremental variance reduction after a further 
prewhitening was less than 1\%}.

{\correction The results of the frequency searches performed unto the 
weighted B- and V-data sets are presented in Tables~\ref{search_freq_A} 
and ~\ref{search_freq_B}. In the former, we list the multi-frequency solution
derived from a weighted combination of two independent analyses: we mention the 
identification number of the frequency, the frequency value (weighted mean), the 
error in frequency (weighted variance of the individual errors computed with {\sc 
Period04}), the signal-to-noise ratio (S/N or significance) and the reduction 
of the relative variance in two filters:
\begin{eqnarray}
 R &=& 1 - (\sigma_{residual}/\sigma_{original})^2.
\end{eqnarray}
In both cases, we achieved a total reduction of the relative variance of the order 
of 80-85\%. 
In the latter, we list the amplitude, the phase, and the residual standard deviation 
as derived from a multi-parameter least-squares fit of sinusoidal functions applied 
to each time series. We further mention the amplitude ratio A$_{B}$/A$_{V}$ (col.~8) and 
the phase difference ($\Phi_{V}$-$\Phi_{B}$) (col.~9) with their respective errors.
When accurate, such observed values may be compared to a theoretically derived amplitude 
ratio and phase difference for a possible determination of the spherical harmonic degree 
$\ell$ of the pulsations at a given pulsation constant 
(see Sect.~\ref{sec:mode}).

This analysis revealed two major frequencies which appear in all our (individual
as well as {\correction merged}) data sets. These frequencies are 19.747 \cd (F1)
and 12.105 \cd (F2), with a frequency ratio of  0.613. 
Their values, amplitudes and phases are well-determined. 
Six minor frequencies are furthermore common to both searches: these are - in 
decreasing order of least-squares amplitude in the filter B - 9.622 \cd (F3), 
14.272 or 15.275 \cd (F4), 7.911 or 8.911 \cd (F5), 8.105 or 9.105 \cd (F7), 
8.698 or 9.868 \cd (F8), and finally, the coupling frequency 31.852 \cd (S12=F1+F2). 
Though some amplitudes are at the level of a few mmags, all of these frequencies 
nevertheless appear with a significant S/N. This is also the case of 31.852 \cd, 
particularly significant in the set 'B All'. The detection of eight common 
frequencies in both time series lends credibility to their secure identification,
except -- in some cases -- for the daily aliasing.}

\subsubsection{Confirmation and interpretation}
\label{sec:confirmation}

{\correction We additionally performed the search for the best multi-frequency solution (in 
the sense of a full scale least-squares analysis) using a different method. The data 
collected in both filters B and V, with their respective weights attributed on a night-to-night 
base, were simultaneously fitted with a same set of frequencies, but with amplitudes and phases 
different for each filter. It was not feasible to explore the complete frequency space in six
or more dimensions, but the region from 8 to 10 \cd was always well covered. A converging 
six-frequency solution was found with F1 = 19.74727 \cd, F2 = 12.10489 \cd, F3 =
9.62244 \cd, F4 = 14.27155 \cd, F5 = 8.91111 \cd and S12 = 31.85223 \cd. These frequencies 
are within the errors identical to the first five frequencies and the sum frequency of the 
adopted solution based on the weighted averages of the individual frequency-analyses of 
the B- and V-time series (cf. Table~\ref{search_freq_A}). The associated amplitudes 
and phases are identical to those of Table~\ref{search_freq_B}.}

{\correction It is interesting to compare our frequency-solution with the one based 
on the {\sc Hipparcos} measurements: although F1 is clearly present in both, the second 
most dominant frequency (F2) is not. Instead, the next possible frequency was found at 
14.268 \cd, which matches extremely well with one of the two probable values for F4.
This is unambiguous evidence for amplitude and/or frequency modulation. Because of
the possible detection of 14.272 \cd (F4) in the {\sc Hipparcos} measurements, we 
gave preference to this frequency in the adopted multi-frequency solution. However,
the alternative choice at 15.275 \cd (F4) would not lead to a different 
frequency-solution than the one listed in Table~\ref{search_freq_A}.}  


{\correction One frequency of the B-time series 
(i.e. F6 = 2.000 \cd) has no matching frequency in the V-time series.  We therefore 
consider this frequency as not being caused by a real effect. In general, we would not 
trust any detection at a frequency below 4 \cd (such as in the case of set 'B All') due 
to the treatment and the combination of a variety of data sets. We furthermore remark that, 
though the B-data are more numerous, we obtained a somewhat smaller significance level for 
the most dominant frequency (F1) with the B-data than with the V-data. However, the opposite 
is true for all the other frequencies.}

\begin{figure*}
  \center
  \resizebox{\textwidth}{!}
  {\includegraphics[angle=0,height=1cm,clip=]{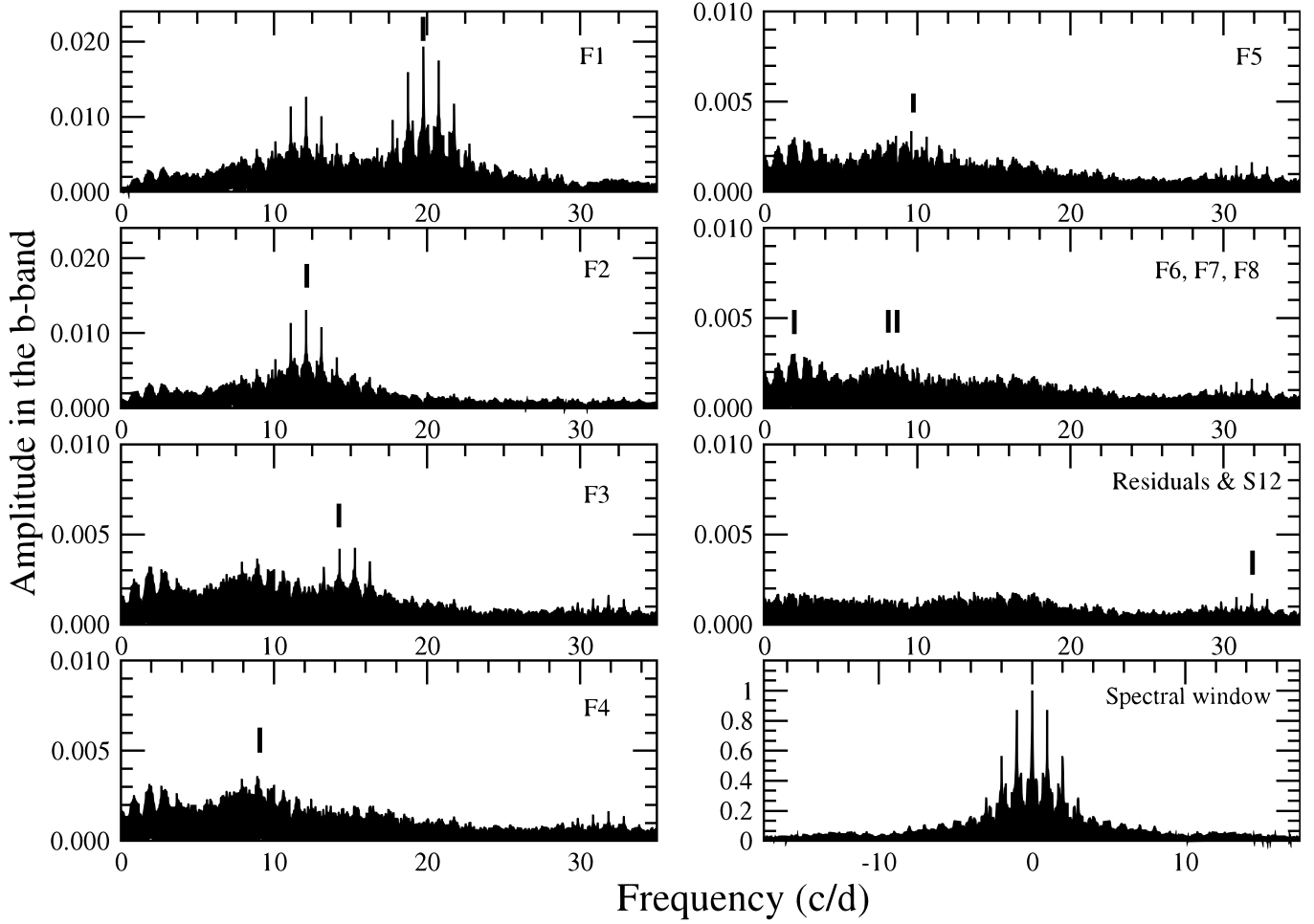}}
    \caption{\correction Successive frequency searches and spectral window of the Fourier
     analysis (Set~All - filter B)\label{fig:periodB}}
 \end{figure*}

\begin{figure*}
  \center
  \resizebox{\textwidth}{!}
  {\includegraphics[angle=0,height=1cm,clip=]{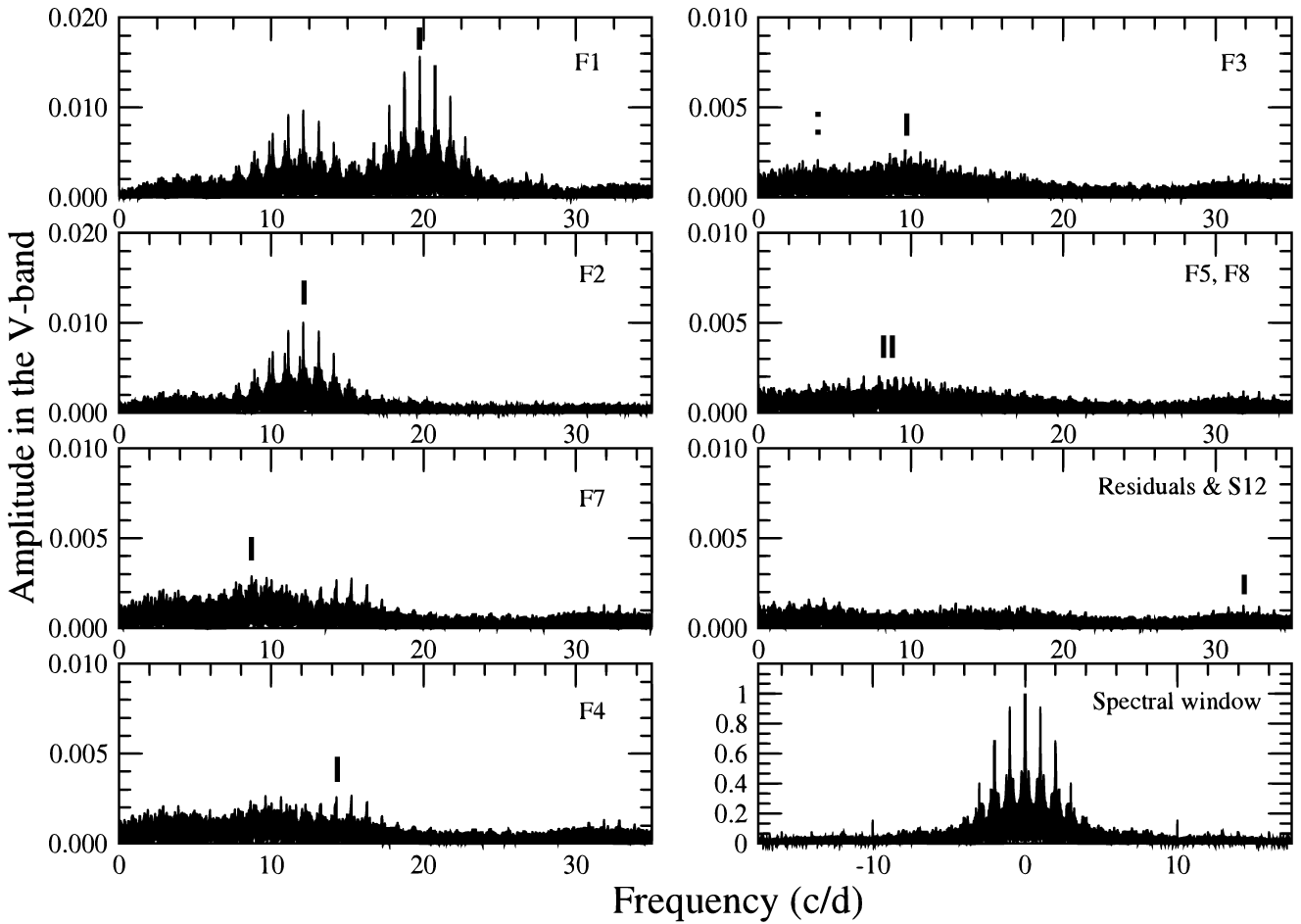}}
    \caption{\correction Successive frequency searches and spectral window of the Fourier
     analysis (Set~All - filter V)\label{fig:periodV}}
 \end{figure*}

Figs.~\ref{fig:periodB} and ~\ref{fig:periodV} illustrate both frequency
searches. We plotted the Fourier spectrum of the original data followed
by those of the residual data after every successive prewhitening. For
sets 'B All' and 'V All', we stopped the search after {\correction the detection 
of 8} frequencies. The {\correction coupling frequency} at 31.852 \cd was not identified as the next 
highest peak in the periodograms of the residual data sets, but the obvious pattern 
imitating the spectral window is a clear indication of its unambiguous identification. 
Since its S/N is {\correction at least equal to 3} in both sets, we also included 
this frequency in the final fitting process.

Two parallel frequency searches performed on the differential magnitudes (K - C1)
revealed no significant peaks with an amplitude larger than 1 mmag in both time series. We
estimated upper limits for the expected noise level from these data (the differential
magnitude is larger than for the variable star): we obtained 9.6 mmag (set 'B All') and 9.0
mmag (set 'V All'). As expected, the final scatters of the residual data appear to be much 
lower: we obtained 7.5 and 7.4 mmag for the B- and the V-residuals respectively (cf. 
Table~\ref{search_freq_B}). We {\correction assume} that these values reflect the 
true noise levels of our {\correction merged} data sets.

\begin{table}[]
\begin{center}
\caption{Adopted frequency-solution based on the weighted analyses of the sets 'B All' and 'V All'
\label{search_freq_A}}
\begin{tabular}{crrrccc}
\hline
\hline
 Freq. & Freq.  & Error  & S/N  & R (\%) & S/N  & R (\%) \struutup\\
 Id.   & {\small (\cd)}  & {\small (\cd)} &\mcol{2}{c}{\it Filter B} &\mcol{2}{c}{\it Filter V} \struutdown\\
\hline
 F1 &     19.747258   &  0.000013  & 41.4  &  51 &  44.6 &  50 \struutup\\
 F2 &	  12.104912   &  0.000019  & 24.5  &  75 &  20.0 &  71 \\
 F3 &	   9.622399   &  0.000078  &  8.3  &  76 &   7.3 &  72 \\
 F4 &	  14.271621$^{a}$  &  0.000059  &  6.9  &  79 &   4.9 &  73 \\
 F5 &	   8.910976$^{a}$  &  0.000083  &  7.6  &  81 &   6.9 &  75 \\
 F6 &	   2.000229$^{b}$  &  0.000075  &  5.6  &  82 &    -- &  -- \\	
 F7 &	   8.69806$^{a}$   &  0.00011   &  5.6  &  83 &   6.3 &  76 \\
 F8 &	   8.10529$^{a}$   &  0.00012   &  4.6  &  84 &   6.2 &  78 \\
 S12 &    31.85217    &  0.00014   &  6.9  &  84.6 & 3.3 &  78.0 \struutdown\\
\hline
\end{tabular}
\end{center}
 $^{a}$: these values may be affected by the 1 \cd (and 0.003 \cd) aliasing i.e. 15.275, 9.698 and 9.105 
 \cd give equally good fits. Preference was given to a solution with 14.272 (since marginally found in the 
 Hipparcos data), 8.911 \cd (peak convincingly detected in 'B All') and 8.698 (peak convincingly detected 
 in 'V All').\\ 
 $^{b}$: this frequency was not found in 'V All', but there is an unsignificant peak at 2.89 or 3.89 \cd.
\end{table}

\begin{table}[]
\begin{center}
\caption{{\correction Amplitude(s) (ratios) and phase(s) (differences) in two filters for the adopted 
frequency-solution. All the phases are computed with respect to the initial epoch of set 'B All'. 
The standard errors on the derived parameters are also shown.}
\label{search_freq_B}}
\begin{tabular}{@{\extracolsep{-2mm}}c@{\extracolsep{0mm}}r@{\extracolsep{0mm}}r@{\extracolsep{-2mm}}r@{\extracolsep{2mm}}r@{\extracolsep{0mm}}r@{\extracolsep{-2mm}}r@{\extracolsep{2mm}}r@{$\pm$}@{\extracolsep{-0.1mm}}l@{\extracolsep{1 mm}}r@{$\pm$}@{\extracolsep{0mm}}l@{\extracolsep{-2mm}}}
\hline
\hline
 Freq. & \mcol{1}{c}{A$_{B}$}  & \mcol{1}{c}{$\sigma_{B}$} & \mcol{1}{c}{$\Phi_{B}$} & \mcol{1}{c}{A$_{V}$} & \mcol{1}{c}{$\sigma_{V}$} & \mcol{1}{c}{$\Phi_{V}$} & \mcol{2}{c}{A$_{B}$/A$_{V}$}  & \mcol{2}{c}{$\Phi_{V}$-$\Phi_{B}$}\struutup\\
 Id. & \mcol{2}{c}{\hspace{1mm}(mmag)} & \hspace{1.5mm}{(2$\pi$rad)} & \mcol{2}{c}{\hspace{1mm}(mmag)} & \hspace{1.5mm}{(2$\pi$rad)}& \mcol{2}{c}{} & \mcol{2}{c}{(2$\pi$rad)} \struutdown\\
\hline
\mcol{1}{c}{}& \mcol{3}{c}{\it Filter B (19.0)} & \mcol{3}{c}{\it Filter V (15.7)} & \mcol{2}{c}{}& \mcol{2}{c}{}\struut\\
\hline
 F1& 19.3  &  13.3   & 0.39  & 16.0 &  11.1 & 0.40  & 1.21 &{\tiny 0.02} & 0.006 &{\tiny 0.003}\struutup\\
 F2& 13.4  &   9.5   & 0.23  & 10.7 &   8.5 & 0.24  & 1.25 &{\tiny 0.04} & 0.008 &{\tiny 0.005}\\
 F3&  4.3  &   9.3   & 0.46  &  2.7 &   8.3 & 0.44  & 1.54 &{\tiny 0.16} & $-$0.01 &{\tiny 0.02}\\
 F4&  4.2  &   8.8   & 0.28  &  2.9 &   8.1 & 0.30  & 1.34 &{\tiny 0.15} & +0.01 &{\tiny 0.02}\\
 F5&  3.9  &   8.3   & 0.60  &  3.2 &   7.9 & 0.59  & 1.35 &{\tiny 0.13} & $-$0.01 &{\tiny 0.02}\\
 F6&  3.6  &   8.0   & 0.54  &  --  &    -- &    -- & \mcol{2}{c}{--} &   \mcol{2}{c}{--} \\
 F7&  2.9  &   7.7   & 0.56  &  2.7 &   7.7 & 0.53  & 1.09 &{\tiny 0.14}  & $-$0.03 &{\tiny 0.02}\\
 F8&  2.4  &   7.6   & 0.87  &  2.7 &   7.4 & 0.91  & 0.91 &{\tiny 0.12}  & +0.04 &{\tiny 0.02}\\
 S12& 1.7  &   7.5   & 0.42  &  1.3 &   7.4 & 0.45  & 1.3  &{\tiny 0.3}  & +0.03 &{\tiny 0.04}\struutdown\\
\hline
\end{tabular}
\end{center}
\end{table}

\subsubsection{Attempt of mode identification}
\label{sec:mode}

{\correction The atmospheric stellar parameters listed in Table~\ref{tab:stellarparameters} were obtained 
through spectroscopic synthesis of high S/N and high-resolution spectra. Therefore, the stellar parameters 
of HD~217860 are known with a relatively good accuracy. This enabled us to compute, together with the 
frequencies of Table~\ref{search_freq_A}, reliable values for the pulsation constant which are useful 
to attempt identification of the excited modes. The following expression can be used here \citep{2000ASPC..210....3B}:

\[log~Q = - log~f + 0.5~log~g + 0.1~M_{bol} + log~T_{eff} - 6.456,\]   

\noindent
allowing a relative accuracy of about 7\%~on the pulsation constant. Deriving M$_{bol}$ = 1.65 $\pm^{0.07}_{0.13}$
from the star's location in the H-R diagram, we obtained the pulsation constants of Table~\ref{tab:pulscon}. 
We can see that the two most dominant modes may correspond to radial (overtone) modes, while 
F3 could possibly be identified as the fundamental radial mode (F). Remark, however, that the corresponding frequency ratio F2/F1 is 0.613 \citep[this value matches the standard ratio expected 
for 2H/F,][]{1997ApJ...477..346B}.
} 

\begin{table}[]
\begin{center}
\caption{Derived pulsation constants for the frequencies of the adopted solution and possibly corresponding mode identification.}
\label{tab:pulscon}
\begin{tabular}{crrc}
\hline
\hline
 Freq. & \mcol{1}{c}{Q}  & \mcol{1}{c}{$\sigma_{Q}$} & Mode \struutup\\
 Id. & \mcol{1}{c}{(days)} & \mcol{1}{c}{(days)} &  Id.\struutdown\\
\hline
 F1& 0.016(3)  &  0.001(1) & poss. 3H\struutup\\
 F2& 0.026(6)  &  0.001(8) & poss. 1H\\
 F3& 0.033(4)  &  0.002(2) & poss. F \\
 F4& 0.022(5)  &  0.001(5) & poss. 2H\\
 F5& 0.036(1)  &  0.002(4) & non-radial? \\
 F7& 0.037(0)  &  0.002(4) & non-radial? \\
 F8& 0.039(7)  &  0.002(6) & non-radial? \struutdown\\
\hline
\end{tabular}
\end{center}
\end{table}

 \begin{figure}
  \center
  \includegraphics[angle=0,height=5cm,clip=]{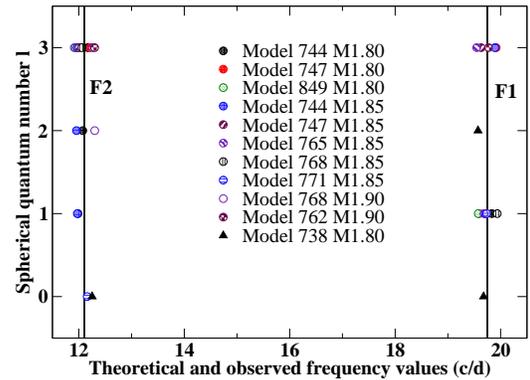}
    \caption{\correction  Theoretically predicted versus observed frequencies for F1 and F2. Vertical lines 
    represent the observed frequency values.\label{fig:plotF12}}
 \end{figure}

\begin{figure*}
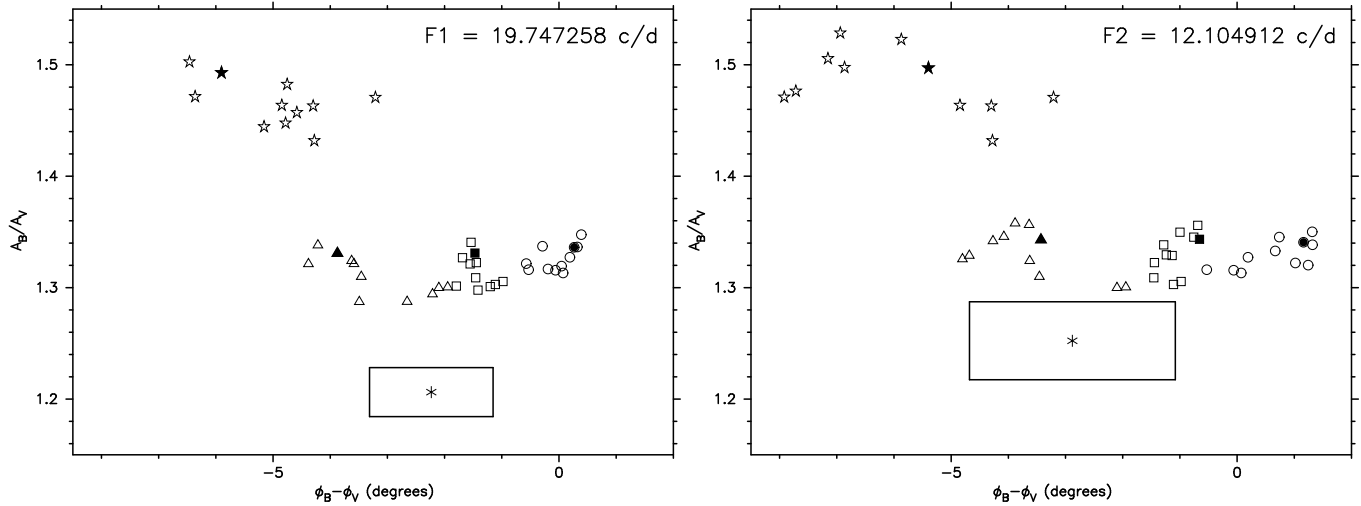

  \center
  \resizebox{\textwidth}{!}
  {\includegraphics[angle=270,clip=]{H113790_IDplot1_final.eps}
  \includegraphics[angle=270,clip=]{H113790_IDplot2_final.eps}}
  \caption{\correction Observed and theoretical values in an (amplitude ratio 
vs. phase difference) diagram for the $\ell$\,= 0 (circles), 1 (squares), 
2 (triangles), and 3 (stars) modes whose frequency is the closest to the 
observed frequencies F1 (left panel) and F2 (right panel) for the 11 
selected models of mass 1.80, 1.85, and 1.90\,M$_{\odot}$. The amplitude 
ratios and the phase differences corresponding to the extra model 
\#738 of mass 1.80\,M$_{\odot}$ correspond to the filled symbols. 
One$\sigma$-error boxes are plotted around the observed values. \label{fig:IDplots}}
 \end{figure*}

{\correction Additionally, we used the most recently developed models for $\delta$ Scuti stars 
including also stellar atmospheres with low-efficiency convection \citep{2007astro.ph..3400M} to 
perform a comparison between the theoretical frequency values of modes with degree $\ell$\,= 0, 
1, 2 or 3 and the observed two most dominant ones. Only for these frequencies, an accurate determination 
of the amplitudes and phases in B and V is possible. We selected 19 appropriate models of mass 1.80, 1.85 
and 1.90 M$_{\odot}$ predicting F1 at a value less than 0.2 c/d from its observed value and whose 
atmospheric stellar parameters satisfy the conditions in effective temperature and gravity 
of $\Delta$~log~\teff $<$ 0.015 and $\Delta$~log~g $<$ 0.1. Then, we looked at those models which 
also predict F2 at less than 0.2 c/d from its observed value: 10 models remained. 
Fig.~\ref{fig:plotF12} illustrates the differences for F1 and F2 versus the spherical 
harmonic degree $\ell$ for the 10 models matching all the requirements. From this comparison, we see 
that most models indicate non-radial pulsation with $\ell$\,= 1 or 3 for F1, and $\ell$\,= 2 or 3 for F2. 
In one such case (model \#771 with a mass of 1.85 M$_{\odot}$ and $\Delta$~log~\teff = 0.012), we found 
a radial mode ($\ell$ = 0, identified as the first overtone 1H) for F2 and a non-radial one ($\ell$ = 1) 
for F1. In one more model, but a slightly less evolved one lying just outside the range of allowed parameters 
(model \#738 with a mass of 1.80 M$_{\odot}$ and $\Delta$~log~g = 0.12), we found excitation of two 
radial modes: the fundamental radial mode (F) for F2 and the second overtone (2H) for F1 (another but poorer match indicated $\ell$\,= 2 for F1) (cf. triangles in Fig~\ref{fig:plotF12}).  

Another tool possibly useful for the determination of the spherical harmonic degree $\ell$ of the pulsations 
at a given value of the pulsation constant \citep{2000ASPC..210...67G, 2003A&A...398..677D} is the comparison between observed 
photometric and theoretical values of the amplitude ratio and the phase difference provided by two-colour 
photometry. We used the same models as in Fig.~\ref{fig:plotF12} to derive the non-adiabatic quantities 
for the $\ell$\,= 0, 1, 2, and 3 modes. In Fig.~\ref{fig:IDplots}, 
we show the resulting amplitude ratio versus phase difference diagramme for the 11 models with frequencies 
matching both F1 and F2.
For each model, the harmonic degree $\ell$\ is represented by a different symbol. The amplitude ratios and 
the phase differences corresponding to the extra model \#738 with a mass of 1.80\, M$_{\odot}$ correspond 
to the filled symbols. 
Although, as we can see, there is a discrepancy with the predicted amplitude ratios (notwithstanding the 
inclusion of thin convective zones), the phase differences are in good agreement. We conclude from this 
diagramme that the observed values (Table~\ref{search_freq_B}) might be compatible with low-degree ($\ell$\,= 0, 1 or 2) modes while $\ell$\,= 3 modes are not probable since the theoretical non-adiabatic observables 
lie too far from their observed counterparts.}

\section{Remarks on other interesting targets}

\label{sec:individual}

\subsection{HD 3743 -- HIP 3165}


The star is the primary of a {visual} binary system {also
known as CCDM 00403+2403~A, forming a common proper motion pair
(CPM) with CCDM 00403+2403~B (= HIP~3163) (angular separation =
16.5~\arcsec, position angle = 205.5\degr~and $\Delta Hp$ = 2.29
mag \nocite{1997yCat.1239....0E}, ESA 1997}). Since the circular entrance
{pupil} of the ELODIE {fiber} is 2~\arcsec~across, {only component
A was observed}. However, the
star showed a bottle-shaped CCF. The differential photometric
analysis showed a shift of 0.078~mag in mean light level between
two consecutive nights. Though the scatter is large (about 0.01 mag) and increases at the 
end of each time series �(due to higher airmass), the {\it comparison minus check star} 
data do not reveal this feature.
We therefore suggest that HIP~3165 is a close binary and that the change in mean level
could be related to ellipsoidal variations with a periodicity of several days.
This behaviour is also confirmed by a simple Fourier analysis of the {\sc HIPPARCOS} epoch
photometric data: a period of $\sim$320~days was derived with an amplitude of 0.02~mag.
The {\sc HIPPARCOS} sampling is, however, very scarce and much shorter
periods are thus possible (and even probable).
For all these reasons we propose HD~3743=HIP~3165 as a new SB2, making CCDM~00403+2403~AB
at a least triple system. However, more differential photometric data are needed in order
to confirm this work hypothesis.

\subsection{HD 3777 -- HIP 3227}

HD~3777 was classified as an Am star by \citet{1964bidelman} and by \citet{1965PASP...77..184C}.
\citet{1970A&AS....1....7B} reported broad and weak metallic lines.
The radial velocity of the star varies with an amplitude of about 40 \kps \citep{1992A&AS...94..479D,
1999A&AS..137..451G}, probably due to binarity, as confirmed by the two peaks detected
in the CCFs. However, the secondary is much fainter than its companion and we were able to fit the spectrum assuming only one component. Since the observed \ion{Ca}{ii} K is
fainter than usual, we hereby confirm that this star is a chemically peculiar star of type Am.

\begin{figure}
\center
\includegraphics[width=8cm,clip=]{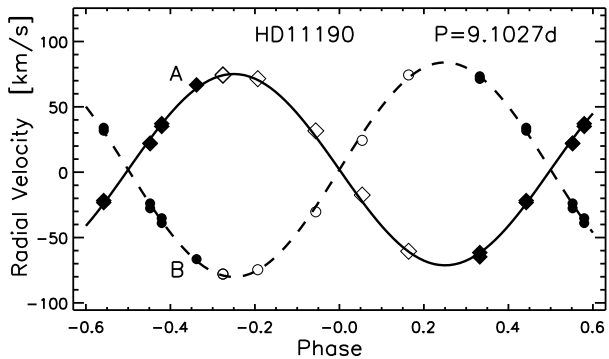}
\caption{Radial velocity phase diagram of HD~11190. Full and dashed lines represent the orbital solution
assuming a circular orbit. Filled symbols correspond to data collected at NAO Rozhen, unfilled ones to
data collected at OHP.\label{fig:vr}}
\end{figure}

\subsection{HD~11190 -- HIP~8581}

Very little is known about HD~11190 except for its variable radial velocity. Actually, this star 
forms a SB2 system consisting of two almost identical components (Fig.~\ref{fig:vr}). {We 
monitored it spectroscopically during five consecutive nights in May 2005 and four nights in 
December 2006}. The orbital parameters we derived from the adjustment of the radial velocity 
measurements assuming a circular orbit are presented in Table~\ref{tab:11190orb}.

\begin{table}[ht]
\caption{Orbital parameters of HD~11190 assuming a circular orbit\label{tab:11190orb}}
\center
\begin{tabular}{r@{}cr@{}c@{}l}
\hline \hline\noalign{\smallskip}
P  &= & 9.1027 &$\pm$& 0.0002 days \\
T$_{\rm 0}$ &=& 2453336.907 &$\pm$& 0.016 JD\\
K$_{\rm A}$ &=& 73.19 &$\pm$& 0.72 \kps\\
M$_{\rm B}$/M$_{\rm A}$ &=& 0.892 &$\pm$& 0.017\\
$\gamma$ &=& 1.93 &$\pm$& 0.35 \kps\\
\hline
\end{tabular}
\end{table}

The fitting of the hydrogen lines provides very similar parameters for the two components and
correspond to an A9~IV-V spectral type. Since the \ion{Ca}{ii} K line of the components is very
weak, and would better agree with that of an A3 star, we may conclude that both components are
Am stars. The luminosity ratio (\lratio~=  2.02$\pm$0.02) and interpolation
through theoretical evolutionary tracks further provide:\smallskip

\begin{tabular}{rcl}
M$_{\rm A}$&=&2.03 $\pm$ 0.19 M$\odot$ \\
R$_{\rm A}$&=&2.97 $\pm$ 0.81 R$\odot$ \\
R$_{\rm A}$/R$_{\rm B}$ &=& 1.39 $\pm$ 0.90.
\end{tabular}
\smallskip

\noindent The dynamical mass ratio is in good agreement with the {luminosity ratio obtained} from
spectroscopy and atmosphere modeling. Although the stellar radii are affected by large error bars,
it is interesting to note that their ratio is almost equivalent to the ratio of the
\vsini~values (1.45$\pm$0.31), which means that the rotation of both components is
probably synchronized with the orbital motion. This is also supported by the low
\vsini~values of both components. Assuming that $i_{spin}$ $\approx$ 90\degr~and
the alignment of the spin and the orbital axes, one would expect values very close
to the observed ones:

V$_{\rm{synch.}}^{\rm{A}}$ = 17 \kps, V$_{\rm{synch.}}^{\rm{B}}$ = 12 \kps.

\noindent Note, however, that {\correction the propagated} error bars {\correction on these estimates} are large (i.e. 29 \kps), which does not allow us to conclude on the spin/orbit inclination itself.

\begin{figure}
\center
\includegraphics[width=8cm,clip=]{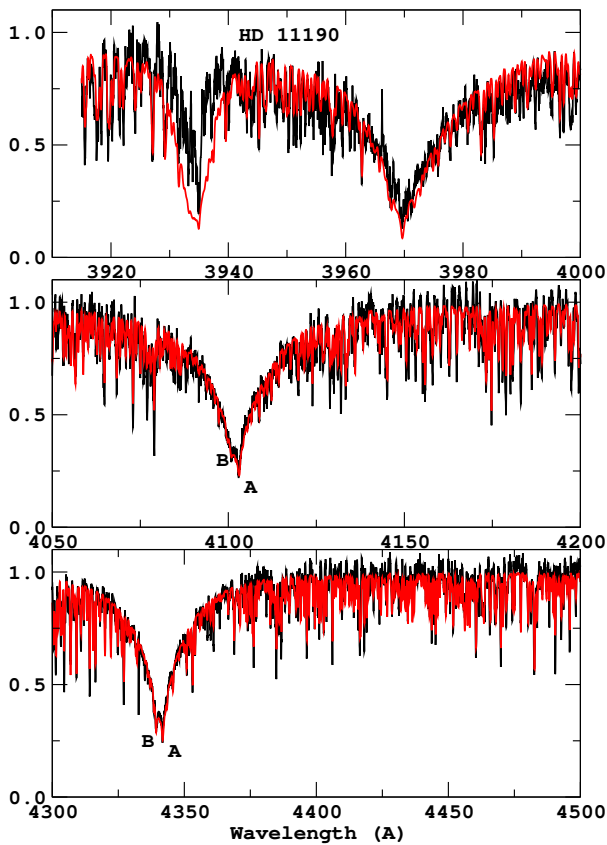}
\caption{Comparison between observed (black line) and
synthetic spectra (red line) in the case of the SB2 system
HD~11190 at HJD=2453343.4033 . The presence of the two components was taken into
account in the fit\label{fig:o4sp}. {\correction The significant disagreement
between observations and model indicates that both components are Am stars.}}
\end{figure}

\subsection{HD~12389 -- HIP~9501}

Colours in the Str\"omgren photometric system are available for this star from \citet{1995IBVS.4216....1H,1999IBVS.4817....1H}.
Adopting the calibration of \citet{1985MNRAS.217..305M}, we found \teff~= 8230 K and \logg~= 3.87 (i.e. corresponds to A5 IV), while
the spectral types provided by SIMBAD and by \citet{1999A&AS..137..451G} are A0 and A4~V,
respectively. The fitting of the ELODIE data with synthetic spectra results in an effective temperature
in good agreement with the one obtained from uvby photometry.
{The CCF shows clear LPVs.} We obtained one light curve on JD~2453359 (cf. Fig.~\ref{light:curves}).
Despite a dip at the beginning of the light curve of the {\it comparison minus check star}, the data are of very good quality. Though the {\it star minus comparison} data show a standard
deviation of at most 6~mmag, the presence of regular, short-period and small-amplitude variations is
obvious. This star is actually a known $\delta$ Scuti variable star
previously detected by \citep{1991AJ....101.2177S}. A period of 0.04~day and a total peak-to-peak
amplitude of 0.03~mag were reported \citep{2000A&AS..144..469R}. From our data, we estimated a similar period
($\sim$ 0.045~day) and amplitude.

\subsection{HD~68725 -- HIP~40361}

HD~68725 was first recognized as a peculiar Am or Ap star by \citet{1980A&AS...39..205O}
on the basis of Str\"omgren photometry, then by \citet{1984ApJ...285..247A} who classified
it as kF2/hF5/mF6 from 1\,\AA~resolution spectra, while \citet{1999A&AS..137..451G} classified
it F2Ib. In the H-R diagram it is located at the cooler edge of the CIS and close to the 
{\correction Terminal Age Main Sequence} (TAMS; Fig.~\ref{fig:hr}), rather than being a 
supergiant star. Chemical peculiarities are found when comparing our data to synthetic spectra. 
They are mainly visible for the rare earth elements:
$\lambda$4078, 4215 Sr {\sc ii} resonance lines are significantly stronger in the observed
spectra while the iron peak elements are only marginally stronger. Scandium and calcium are
nearly solar. HD~68725 is therefore marginally an Am star, but the enhanced Sr possibly
indicates the presence of a magnetic field. 
{\correction In their survey of the Solar neighbourhood, \citet{2004A&A...418..989N} reported 
[Fe/H]=$+0.33$ for this star.} \citet{2001AJ....121.3224M} observed it twice
using speckle-interferometry but found no evidence for multiplicity. However, the night-averaged 
CCF shows a slight {and systematic} bump on the left side of the main peak which could be related 
to the presence of a secondary component. This seems to be confirmed when comparing the averaged 
radial velocity we obtained (RV = $-$6.9 $\pm$ 4.1 \kps) to the ones previously published by 
\citet[][$-$36 $\pm$ 0.8 \kps]{1990A&AS...83..251D}, \citet[][$-$17.5 $\pm$ 5.6 \kps]{1999A&AS..137..451G} 
and \citet[][$-$10.9 $\pm$ 1.1 \kps]{2004A&A...418..989N}. {\correction It is worth mentioning that, 
we cannot exclude at the present time that this bump could also be due to a spot, which could 
provide another proof of an existing magnetic field}.

The CCF of HD~68725 further shows rapid shape changes (Fig.~\ref{fig:ccf}), probably
caused by pulsation of the $\delta$ Scuti-type as we indeed detected short-period variations
in the CCD photometric data. The light curve in Fig.~\ref{light:curves} suggests a period of
$\sim$0.12 days. HD~68725 is therefore a chemically peculiar star which also does pulsate as a typical $\delta$ Scuti star. More spectra are however needed to decide whether it is
a mild Am or a Ap star. {If it would prove to be a magnetic Ap star (showing magnetic variations along with spectral variations modulated by the rotational period), it would be an
exceptional case, together with HD~75425, a weak Ap Sr(CrEu) star \citep{1998Ap&SS.259...57M} and
HD~188136, a $\rho$~Pup star \citep{1980MNRAS.193...51K}. Such cases suggest that the presence
of a global magnetic field does not inhibit $\delta$ Scuti pulsation in all cases \citep[in contradiction with the actual knowledge of the diffusion theory, see][]{2000BaltA...9..253K}

\subsection{HD~81995 -- HIP~46642}

The HIPPARCOS Epoch photometry of HD~81995 varies with time
in an unclear way (unsolved HIPPARCOS variable). The CCF is affected by short-term
LPVs and by radial velocity variations due to the presence of a much fainter companion. The star
is therefore SB1. We additionally obtained two light curves on JD~2453462 and JD~2453471. Both
time series show a partial eclipse, with a drop in magnitude of about 0.06 mag. The standard
deviations on the {\it comparison minus check star} data are 5 and 7 mmag respectively.
The light curves are typical of a detached or semi-detached binary (Fig.~\ref{light:curves}).
A third light curve obtained on JD~2453472 showed no changes and a standard deviation of only
4 mmag. In the latter time series, the comparison star appeared to be unreliable.


\subsection{HD~221774 -- HIP~116321}

HD~221774 was found to be a double-lined spectroscopic binary in which the primary and
the secondary components have spectral types A6 and F1, respectively. Small changes
were detected in the high S/N CCFs of this binary. We therefore observed it photometrically for two
partial nights. Both {\it star minus� comparison} time series have a standard deviation of 4 mmag
and do not show any detectable sign of short-period variability, while the standard deviation of {\it comparison minus check star} data was 6 mmag. The light curve in Fig.~\ref{light:curves}
illustrates the data obtained on JD~2453408. We thus classified HD~221774 as a photometrically
constant star.


\begin{figure}
\center
\includegraphics[width=8.5cm,angle=0,clip=]{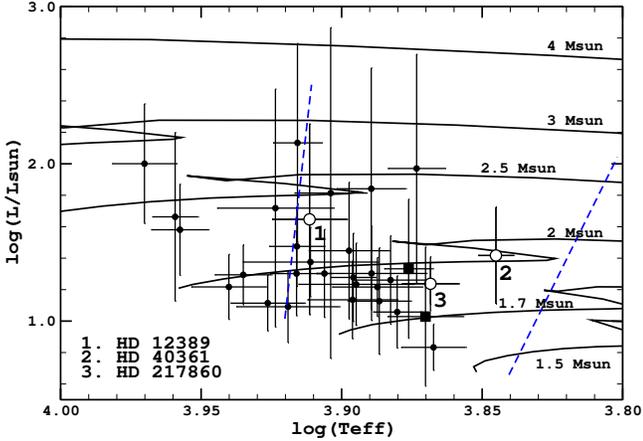}
\caption{Location of the targets in the H-R diagram. Open circles
are the stars we identified as short-term variables of $\delta$
Scuti type. Squares stand for the components of the SB2 binary
HD~11190. Other SB2 systems with {inaccurate} parameters were not
plotted here. Broken lines are the limits of the Cepheid
instability strip as computed by \citet{2005A&A...435..927D}
\label{fig:hr}}
\end{figure}

\begin{table}
\center
\caption{List of interesting targets\label{tab:intarg}}
\begin{tabular}{rl}
\hline \hline\noalign{\smallskip}
HIP & Remarks\\
\hline\noalign{\smallskip}
\multicolumn{2}{c}{Intrinsically variable stars}\\ \noalign{\smallskip}
\hline\noalign{\smallskip}
9501 & known $\delta$ Scuti\\
40361 & SB2? or/and CP/Am star\\
113790 & $\delta$ Scuti, multiperiodic\\
\hline\noalign{\smallskip}
\multicolumn{2}{c}{Spectroscopic systems}\\ \noalign{\smallskip}
\hline\noalign{\smallskip}
3165 & SB2, ellipsoidal?\\
3227 & SB2, Am star\\
5416 & SB2\\
8581 & SB2\\
9851 & SB1?\\
40361 & SB2, CP/Am star\\
81995 & SB1, eclipsing\\
116321 & SB2\\
\hline
\end{tabular}
\end{table}

\section{Discussion and conclusions}

\label{sec:conclusions}

We obtained several high-resolution, high signal-to-noise spectra distributed over various 
time scales ranging from a couple of hours to some days for a sample of 32 bright though
poorly investigated A-type stars
selected on the basis of their variable radial velocity \citep{1999A&AS..137..451G}.
These spectra were supplemented by differential CCD light curves collected as time series
during parts of at least two nights in seven cases, in order to look for rapid light
variations. Among the 32 investigated targets, we discovered {\correction eight spectroscopic 
binaries, one of which is a close (photometric) binary (HD~81995)}. 
In one case, we claim the detection of ellipsoidal variability caused by the proximity
of both components (HD~3165). In all the cases, the components have projected rotational
velocities below 100 \kps, which can be expected if spin-orbit synchronization already occurred.
We suggest this state explicitly in the case of the newly detected SB2 (HD~11190), a
twin system consisting of two similar components orbiting around one another with a period
of 9.1~days.If we except HD~13162, HD~25021 (Am star), HD~38771 and HD~217860 (a pulsating
star), all remaining targets have large \vsini~values associated to broad CCFs (duplicity or
multiplicity is not easily detectable in these stars, therefore the high scatter in the previous
radial velocity measurements can be misleading and should not be interpreted
as an indication of variability or multiplicity 'per se'). 

{\correction Among the 32 investigated targets, we also discovered three small-amplitude 
$\delta$~Scuti pulsators showing short-term line-profile variations (LPVs),} an outcome 
confirmed by their CCD differential light curves: one is the already known $\delta$ Scuti 
star HD~12389 {\correction \citep{1991AJ....101.2177S}}, the other two are newly identified 
$\delta$~Scuti variable stars (HD~68725 and HD~217860). The former one (HD~68725) is also a 
member of a marginally metal-enhanced spectroscopic binary showing strong strontium absorption 
lines, making it {\correction the most intriguing system of our sample}. 


{Because both the spectra and the light curves of the pulsating star HD~217860 displayed a highly 
multiperiodic behaviour, we organised an {\correction extensive} photometric campaign for this object 
during the winter of 2005. Observations using the {\correction Bessell} filters B and V were gathered 
at three different sites. The {\correction independent} analyses of two data sets (the sets 'B All' 
and 'V All') enabled us to identify {\correction eight frequencies} common to both time series. 
{\correction The adopted solution was confirmed by a simultaneous multi-frequency search of the 
joint B- and V-data sets.} The two most dominant frequencies detected show a ratio of 0.61; {\correction 
this accurately determined ratio matches extremely well the standard value expected for 2H/F. 
However, this ratio appears to be also compatible with the computation of the respective pulsation 
constants (Q$_{1}$ = 0.016 and Q$_{2}$ = 0.027 days), which rather indicate 3H/1H (with a theoretical 
period ratio of 0.60). As a next step, we computed theoretical versus observed frequency differences 
using 19 appropriate models of mass 1.80, 1.85 and 1.90~M$_{\odot}$, including convection modelling in 
the outer layers \citep{2007astro.ph..3400M}. We performed a comparison between the observed amplitude 
ratios and phase differences for F1 and F2 on the one hand and their theoretical counterparts on the
other hand for 11 models predicting frequencies close to F1 and F2 for $\ell$\,= 0, 1, 2, or 
3 modes. 
From this amplitude ratio versus phase difference diagramme, we concluded that non-radial modes with 
$\ell$\,= 3 are not probable. 
We also showed that most models indicate non-radial pulsation modes with $\ell$\,= 1 for F1, and $\ell$\,= 2 for F2. 
However, there exists one model satisfying the derived stellar parameters for HD~217860 and predicting a radial mode for F2 (as the first overtone, 1H). In addition, we found
a less evolved 1.80 M$_{\odot}$ model (lying just outside of the range of allowed stellar parameters with 
$\Delta$ log~g = 0.12), which is consistent with the conclusion that both frequencies might correspond 
to radial  modes, with the second overtone ($\ell$\,= 0, 2H) and the fundamental ($\ell$\,= 0, F) modes 
being excited. Our conclusion is that some of the models in the right location of the H-R diagram predict pulsation in at least one radial overtone mode among the two major frequencies. 
Among the remaining frequencies with almost equally small amplitudes, the presence of non-radial (mixed) modes is very probable, but we cannot exclude that another radial mode may co-exist (e.g. the frequency F3 with Q$_{3}$ = 0.033 days).}  


We also detected - though at the limit of the allowed significance level - a high-frequency 
component at 31.852 \cd {\correction (or period of about 45 min), corresponding to the coupling 
frequency F1+F2.} 

{\correction More importantly perhaps, we found clear evidence for a strong modulation of the 
amplitude(s) and/or (radial) frequency content in this star since a frequency-analysis of the 
{\sc Hipparcos} Epoch Photometry did not reveal the presence of the frequency that we could 
identify as the possible first radial overtone (F2). It is possible that this $\delta$ Scuti 
star is a radial overtone pulsator with a variable (radial) modal content.
For this reason, it is an important case study for the comparison with models of pulsational stability in the middle of the $\delta$ Scuti instability strip \citep{1997ApJ...477..346B}.
Another, recently investigated, multiperiodic $\delta$ Scuti star shows a very similar behaviour: AN Lyn is an atypical $\delta$ Scuti star showing peculiar and highly multiperiodic light curves such as HD~217860, for which three independent frequencies (10.1756, 18.1309 and 9.5598 \cd)
were determined \citep{1997A&A...328..235R}. Changes in amplitude on a long time scale are 
confirmed for its main frequency \citep{2002A&A...385..503Z}. Furthermore, \citet{2007astro.ph..3400M}
concluded on the basis of a comparison between observed and theoretical values of the amplitude 
ratio and the phase difference that radial (overtone) modes are most probably excited. } 

Given the circumstances and the data, we cannot extract more information for this target
at present, but the present results amply illustrate that new high-quality time series of
differential data organised at a later date would be very worthwhile in order {\correction to 
accurately determine the full frequency content as well as its time-dependent behaviour.
A follow-up spectroscopic campaign would be needed to put additional contraints on the mode identification 
and to enable an in-depth study of the pulsations of this new and atypical $\delta$~Scuti pulsator.}}

{\correction Our future work will be} to analyse and further exploit the high-resolution spectra
obtained here in the context of a {\correction broad} study of the chemical composition of (poorly 
studied) main-sequence stars located in the lower end of the Cepheid instability strip, with the main 
objective to investigate the possible connection(s) between pulsation, radiative diffusion and 
multiplicity in this {\correction intriguing} part of the H-R diagram.

\begin{acknowledgements}
We acknowledge funding from the Belgian Federal Science Policy in the framework 
of the projects "Modern Aspects of Theoretical and Observational (ground-based 
and space-borne) Astrophysics" (Ref. IAP P5/36) and "Pulsation, chemical composition 
and multiplicity in main-sequence A- and F-type stars" (Ref. MO/33/018). {\correction 
DD is grateful to Dr. Z. Kraicheva for many useful discussions and suggestions }. 
This research is based on data obtained at the {\it Observatoire de Haute-Provence}, 
the {\it Observatoire du Pic du Midi} (France), the Hoher List Observatory (Germany) 
as well as the NAO Rozhen, operated by the Institute of Astronomy, Bulgarian Academy 
of Sciences (Bulgaria). {\correction We acknowledge the support of the Belgian 
Science Policy and the Bulgarian Academy of Sciences via bilateral project 
"Photometric and spectroscopic follow-up studies of binary systems of special 
interest"}. The spectroscopic campaigns were funded by the Optical Infrared 
Coordination network (OPTICON) supported by the Research Infrastructures Programme 
of the European Commission. We kindly thank Drs. Hubeny and Lanz for making their 
computer codes available. We furthermore thank Dr. K. Reif, director of the Hoher List 
Observatory from the AIfA (Argelander Institute for Astronomy, University of Bonn), 
for the allocated telescope time. Part of the photometric data were acquired with 
equipment purchased thanks to a research fund financed by the Belgian National Lottery 
(1999). {\correction The referees made various useful suggestions which are gratefully
appreciated}. This research also made use of the SIMBAD database, operated at the 
{\it Centre de Donn\'ees Stellaires} (Strasbourg, France).
\end{acknowledgements}

\bibliographystyle{aa}
\bibliography{hints}

\begin{thebibliography}{44}
\expandafter\ifx\csname natexlab\endcsname\relax\def\natexlab#1{#1}\fi
\expandafter\ifx\csname url\endcsname\relax
  \def\url#1{{\tt #1}}\fi
\expandafter\ifx\csname urlprefix\endcsname\relax\def\urlprefix{URL }\fi

\bibitem[{{Abt}(1984)}]{1984ApJ...285..247A}
{Abt} H.A., Oct. 1984, \apj, 285, 247

\bibitem[{{Baranne} et~al.(1996){Baranne}, {Queloz}, {Mayor}
  et~al.}]{1996A&AS..119..373B}
{Baranne} A., {Queloz} D., {Mayor} M., et~al., Oct. 1996, \aaps, 119, 373

\bibitem[{{Berry} \& {Burnell}(2005)}]{2005haip.book.....B}
{Berry} R., {Burnell} J., 2005, {The handbook of astronomical image
  processing}, The handbook of astronomical image processing, 2nd ed., by
  R.~Berry and J.~Burnell.~ xxviii, 684 p., 1 CD-ROM (incl.~Astronomical Image
  Processing Software AIP4WIN, v.2.0).~ Richmond, VA: Willmann-Bell, 2005

\bibitem[{{Bertaud}(1970)}]{1970A&AS....1....7B}
{Bertaud} C., Jan. 1970, \aaps, 1, 7

\bibitem[{{Bessell}(1995)}]{1995CCDA....2...20B}
{Bessell} M.S., 1995, CCD Astronomy, 2, 20

\bibitem[{{Bidelman}(1964)}]{1964bidelman}
{Bidelman} W., 1964, ONR Symposium held at Flagstaff, Arizona (see reference in
  Cowley \& Cowley 1965)

\bibitem[{{Bono} et~al.(1997){Bono}, {Caputo}, {Cassisi}
  et~al.}]{1997ApJ...477..346B}
{Bono} G., {Caputo} F., {Cassisi} S., et~al., Mar. 1997, \apj, 477, 346

\bibitem[{{Breger}(2000)}]{2000ASPC..210....3B}
{Breger} M., 2000, In: {Breger} M., {Montgomery} M. (eds.) Delta Scuti and
  Related Stars, vol. 210 of Astronomical Society of the Pacific Conference
  Series, 3--+

\bibitem[{{Castelli} \& {Kurucz}(2003)}]{2003IAUS..210P.A20C}
{Castelli} F., {Kurucz} R.L., 2003, In: IAU Symposium, 20P

\bibitem[{{Cowley} \& {Cowley}(1965)}]{1965PASP...77..184C}
{Cowley} A.P., {Cowley} C.R., Jun. 1965, \pasp, 77, 184

\bibitem[{{Donati} et~al.(1997){Donati}, {Semel}, {Carter}, {Rees}, \& {Collier
  Cameron}}]{1997MNRAS.291..658D}
{Donati} J.F., {Semel} M., {Carter} B.D., {Rees} D.E., {Collier Cameron} A.,
  Nov. 1997, \mnras, 291, 658

\bibitem[{{Donati} et~al.(1999){Donati}, {Catala}, {Wade}
  et~al.}]{1999A&AS..134..149D}
{Donati} J.F., {Catala} C., {Wade} G.A., et~al., Jan. 1999, \aaps, 134, 149

\bibitem[{{Duflot} et~al.(1990){Duflot}, {Mannone}, {Genty}, \&
  {Fehrenbach}}]{1990A&AS...83..251D}
{Duflot} M., {Mannone} C., {Genty} V., {Fehrenbach} C., May 1990, \aaps, 83,
  251

\bibitem[{{Duflot} et~al.(1992){Duflot}, {Fehrenbach}, {Mannone}, {Burnage}, \&
  {Genty}}]{1992A&AS...94..479D}
{Duflot} M., {Fehrenbach} C., {Mannone} C., {Burnage} R., {Genty} V., Sep.
  1992, \aaps, 94, 479

\bibitem[{{Dupret} et~al.(2003){Dupret}, {De Ridder}, {De Cat}
  et~al.}]{2003A&A...398..677D}
{Dupret} M.A., {De Ridder} J., {De Cat} P., et~al., Feb. 2003, \aap, 398, 677

\bibitem[{{Dupret} et~al.(2005){Dupret}, {Grigahc{\`e}ne}, {Garrido},
  {Gabriel}, \& {Scuflaire}}]{2005A&A...435..927D}
{Dupret} M.A., {Grigahc{\`e}ne} A., {Garrido} R., {Gabriel} M., {Scuflaire} R.,
  Jun. 2005, \aap, 435, 927

\bibitem[{{Dworetsky}(2004)}]{2004IAUS..224..499D}
{Dworetsky} M.M., Dec. 2004, In: The A-Star Puzzle, IAU Symposium, No. 224,
  499--504

\bibitem[{{Erspamer} \& {North}(2002)}]{2002A&A...383..227E}
{Erspamer} D., {North} P., Jan. 2002, \aap, 383, 227

\bibitem[{{Erspamer} \& {North}(2003)}]{2003A&A...398.1121E}
{Erspamer} D., {North} P., Feb. 2003, \aap, 398, 1121

\bibitem[{{ESA}(1997)}]{1997yCat.1239....0E}
{ESA}, Feb. 1997, {\it The Hipparcos and Tycho Catalogues}, ESA SP-1200

\bibitem[{{Fr\'emat} et~al.(2005{\natexlab{a}}){Fr\'emat}, {Lampens}, {Van
  Cauteren}, \& {Robertson}}]{2005CoAst.146....6F}
{Fr\'emat} Y., {Lampens} P., {Van Cauteren} P., {Robertson} C.W., Jun.
  2005{\natexlab{a}}, Communications in Asteroseismology, 146, 6

\bibitem[{{Fr\'emat} et~al.(2005{\natexlab{b}}){Fr\'emat}, {Neiner}, {Hubert}
  et~al.}]{2005astro.ph..9336F}
{Fr\'emat} Y., {Neiner} C., {Hubert} A.M., et~al., Sep. 2005{\natexlab{b}},
  A\&A, in press, arXiv:astro-ph/0509336

\bibitem[{{Fr{\'e}mat} et~al.(2006){Fr{\'e}mat}, {Antonova}, {Damerdji}
  et~al.}]{2006CoAst.148...77F}
{Fr{\'e}mat} Y., {Antonova} A., {Damerdji} Y., et~al., Dec. 2006,
  Communications in Asteroseismology, 148, 77

\bibitem[{{Garrido}(2000)}]{2000ASPC..210...67G}
{Garrido} R., 2000, In: {Breger} M., {Montgomery} M. (eds.) Delta Scuti and
  Related Stars, vol. 210 of Astronomical Society of the Pacific Conference
  Series, 67--+

\bibitem[{{Gray}(2005)}]{spectrum}
{Gray} R., 2005, {\\http://www.phys.appstate.edu/spectrum/spectrum.html}

\bibitem[{{Grenier} et~al.(1999){Grenier}, {Baylac}, {Rolland}
  et~al.}]{1999A&AS..137..451G}
{Grenier} S., {Baylac} M.O., {Rolland} L., et~al., Jun. 1999, \aaps, 137, 451

\bibitem[{{Handler}(1995)}]{1995IBVS.4216....1H}
{Handler} G., Jun. 1995, Information Bulletin on Variable Stars, 4216, 1

\bibitem[{{Handler}(1999)}]{1999IBVS.4817....1H}
{Handler} G., Dec. 1999, Information Bulletin on Variable Stars, 4817, 1

\bibitem[{{Hubeny} \& {Lanz}(1995)}]{1995ApJ...439..875H}
{Hubeny} I., {Lanz} T., Feb. 1995, \apj, 439, 875

\bibitem[{{Kurtz}(1980)}]{1980MNRAS.193...51K}
{Kurtz} D.W., Oct. 1980, \mnras, 193, 51

\bibitem[{{Kurtz} \& {Martinez}(2000)}]{2000BaltA...9..253K}
{Kurtz} D.W., {Martinez} P., 2000, Baltic Astronomy, 9, 253

\bibitem[{{Lenz} \& {Breger}(2005)}]{2005CoAst.146...53L}
{Lenz} P., {Breger} M., Jun. 2005, Communications in Asteroseismology, 146, 53

\bibitem[{{Martinez} \& {Medupe}(1998)}]{1998Ap&SS.259...57M}
{Martinez} P., {Medupe} R., 1998, \apss, 259, 57

\bibitem[{{Mason} et~al.(2001){Mason}, {Hartkopf}, {Holdenried}, \&
  {Rafferty}}]{2001AJ....121.3224M}
{Mason} B.D., {Hartkopf} W.I., {Holdenried} E.R., {Rafferty} T.J., Jun. 2001,
  \aj, 121, 3224

\bibitem[{{Montalb\'an} \& {Dupret}(2007)}]{2007astro.ph..3400M}
{Montalb\'an} J., {Dupret} M.., Mar. 2007, ArXiv Astrophysics e-prints

\bibitem[{{Moon} \& {Dworetsky}(1985)}]{1985MNRAS.217..305M}
{Moon} T.T., {Dworetsky} M.M., Nov. 1985, \mnras, 217, 305

\bibitem[{{Neiner} et~al.(2003){Neiner}, {Henrichs}, {Floquet}
  et~al.}]{2003A&A...411..565N}
{Neiner} C., {Henrichs} H.F., {Floquet} M., et~al., Dec. 2003, \aap, 411, 565

\bibitem[{{Nordstr{\"o}m} et~al.(2004){Nordstr{\"o}m}, {Mayor}, {Andersen}
  et~al.}]{2004A&A...418..989N}
{Nordstr{\"o}m} B., {Mayor} M., {Andersen} J., et~al., May 2004, \aap, 418, 989

\bibitem[{{Olsen}(1980)}]{1980A&AS...39..205O}
{Olsen} E.H., Feb. 1980, \aaps, 39, 205

\bibitem[{{Rodr\'{\i}guez} et~al.(1997){Rodr\'{\i}guez}, {Gonz\'alez-Bedolla},
  {Rolland} et~al.}]{1997A&A...328..235R}
{Rodr\'{\i}guez} E., {Gonz\'alez-Bedolla} S.F., {Rolland} A., et~al., Dec.
  1997, \aap, 328, 235

\bibitem[{{Rodr{\'{\i}}guez} et~al.(2000){Rodr{\'{\i}}guez},
  {L{\'o}pez-Gonz{\'a}lez}, \& {L{\'o}pez de Coca}}]{2000A&AS..144..469R}
{Rodr{\'{\i}}guez} E., {L{\'o}pez-Gonz{\'a}lez} M.J., {L{\'o}pez de Coca} P.,
  Jun. 2000, \aaps, 144, 469

\bibitem[{{Schaller} et~al.(1992){Schaller}, {Schaerer}, {Meynet}, \&
  {Maeder}}]{1992A&AS...96..269S}
{Schaller} G., {Schaerer} D., {Meynet} G., {Maeder} A., Dec. 1992, \aaps, 96,
  269

\bibitem[{{Schutt}(1991)}]{1991AJ....101.2177S}
{Schutt} R.L., Jun. 1991, \aj, 101, 2177

\bibitem[{{Zhou}(2002)}]{2002A&A...385..503Z}
{Zhou} A.Y., Apr. 2002, \aap, 385, 503

\end{thebibliography}

\newpage

\Online

\onecolumn


\newpage

\setcounter{table}{1}

\begin{center}

\tablehead{\hline\hline
HD      && HJD$-$2400000 & S/N & exp. & \multicolumn{3}{c}{RV} &Obs.\\
          &&             &   & [s] & \multicolumn{3}{c}{[\kps]} &\\
\hline\noalign{\smallskip}}

\tabletail{%
\multicolumn{9}{r}{\small\sl ... continued on next page}\\
\hline}
\tablelasttail{\hline}
\topcaption{Journal of spectroscopic observations at NAO, OHP and TBL.}

\begin{supertabular}{r@{}rlrrr@{}r@{}rr}
    849 && 53346.2338 & 137 &  1200 & 10.46&$\pm$&5.93 & OHP\\
        && 53346.2493 & 143 &  1200 & 5.17&$\pm$&4.10 & OHP\\
   3066 && 53343.3515 &  59 &   644 & $-$12.15&$\pm$&1.35 & OHP\\
        && 53343.3818 &  70 &  1200& $-$1.75&$\pm$&2.86 & OHP\\
       & & 53344.3245 & 104 &  1200 & 2.68&$\pm$&4.17 & OHP\\
        && 53344.3399 & 104 &  1200 & $-$8.86&$\pm$&2.30 & OHP\\
   3743 && 53346.2782 & 136 &  1200 & $-$8.10&$\pm$&0.69 & OHP\\
        && 53346.2937 & 135 &  1200 & $-$8.68&$\pm$&1.73 & OHP\\
        && 53346.3092 & 136 &  1200 & $-$8.10&$\pm$&0.27 & OHP\\
       && 53723.2489 & 60  & 400   & $-$10.07&$\pm$&5.73 & TBL\\
       && 53723.2541 & 57  & 400   & $-$11.04&$\pm$&0.61 & TBL\\
       && 53723.2593 & 71  & 400   & $-$14.07&$\pm$&1.28 & TBL\\
       && 53723.2645 & 74  & 400   & $-$10.11&$\pm$&1.26 & TBL\\
    3777&A & 53346.3278 & 120 &  1200 & 34.12&$\pm$&0.21 & OHP\\
          &B &                   &        &          & $-$102.83&$\pm$&0.67 & \\
        &A & 53346.3433 & 105 &  1200 & 33.97&$\pm$&0.35 & OHP\\
          &B &                   &        &          & $-$102.13&$\pm$&0.67 & \\
   5066 && 53344.3869 & 111 &   900 & $-$9.39&$\pm$&4.46 & OHP\\
        && 53344.3988 & 108 &   900 & $-$11.11&$\pm$&2.76 & OHP\\
   6813&A & 53346.3709 &  86 &  1200 & 12.33&$\pm$&0.50 & OHP\\
           &B &   &   &    & 3.83&$\pm$&0.50 & \\
        &A& 53346.3864 &  79 &  1200 & 12.32&$\pm$&0.5 & OHP\\
            &  B &   &   &    & 3.27&$\pm$&0.50 & \\
        &A& 53723.2745 & 65  & 400  & 13.95&$\pm$&0.40 & TBL\\
            &  B &   &   &    & 6.22&$\pm$&0.87 & \\
    &A& 53723.2797 & 52  & 400  & 13.55&$\pm$&0.42 & TBL\\
            &  B &   &   &    & 5.28&$\pm$&1.90  & \\
    &A& 53723.2849 & 60  & 400  & 13.57&$\pm$&0.51 & TBL\\
            &  B &   &   &    & 5.23&$\pm$&0.24  & \\
        &A& 53723.2902 & 55  & 400  & 14.36&$\pm$&0.38  & TBL\\
            &  B &   &   &    & 6.76&$\pm$&0.41 & \\
   7551 && 53344.3577 &  84 &   720 & 10.32&$\pm$&4.34 & OHP\\
        && 53344.3676 &  73 &   720 & 12.29&$\pm$&5.23 & OHP\\
   11190&A & 53343.4033 &  79 &  1200 & 74.10 &$\pm$& 0.50  & OHP\\
  &B &                   &       &           & $-$77.91 &$\pm$& 0.48   &  \\
  &A & 53343.4188 &  77 &  1200 & 74.55 &$\pm$& 0.62  & OHP\\
  &B &                   &       &           & $-$77.99 &$\pm$& 0.48   & \\
  &A & 53344.2401 &  75 &  1200 & 71.54 &$\pm$& 0.71  & OHP\\
  &B &                   &       &           & $-$74.64 &$\pm$& 0.63    & \\
  &A & 53345.4048 &  71 &  1200 & 31.61 &$\pm$& 0.45  & OHP\\
  &B &                   &       &           & $-$30.32 &$\pm$& 0.20  &\\
  &A & 53346.4068 &  84 &  1200 & $-$17.51 &$\pm$& 0.53  & OHP\\
  &B &                   &       &           &  24.37 &$\pm$& 0.43 & \\
  &A & 53347.5432 &  28 &  1200 & $-$60.52 &$\pm$& 0.35  & OHP\\
  &B &                   &       &           & 74.41 &$\pm$& 0.35 & \\
  &A & 54077.2242 &  161  & 900 & $-$61.53 & &1.15& NAO \\
  &B &            &   &   & 73.39 &&3.07& \\
  &A & 54077.2397 & 124 & 1200 & $-$64.77&&1.28& NAO \\
  &B & & & & 71.35&&3.03&  \\
  &A & 54078.2333& 215 & 1200 &$-$21.72&& 2.88& NAO \\
  &B && & & 34.00&&4.25&  \\
  &A &54078.2518 & 149 & 1500 &$-$23.25&&1.34& NAO \\
  &B && & &31.73&&3.95& \\
  &A &54079.2205& 211 & 1200 &21.83&&2.43& NAO \\
  &B && & &$-$23.88&&4.15&  \\
  &A &54079.2366& 134 & 1200 &22.23&&2.20& NAO \\
  &B && & &$-$27.59&&4.61&  \\
  &A &54079.4801& 172 & 1200 &37.09&&2.81& NAO \\
  &B && & &$-$35.21&&5.19&  \\
  &A &54079.4959& 93 & 1200 &35.05&&2.42& NAO \\
  &B && & &$-$38.94&&4.27&  \\
  &A &54080.2300& 50 & 1200 &66.77&&2.76& NAO \\
  &B && & &$-$66.53&&5.03&  \\
  12389 && 53346.4230 &  82 &  1200 & -37.54&$\pm$&1.06 & OHP\\
        && 53346.4384 &  81 &  1200 & -38.32&$\pm$&1.70 & OHP\\
  12868 && 53343.4364 &  92 &  1200 & -3.98&$\pm$&0.13 & OHP\\
        && 53343.4518 & 104 &  1200 & -3.85&$\pm$&0.09 & OHP\\
  13162 && 53346.4544 &  68 &  1200 & $-$5.53&$\pm$&0.33 & OHP\\
        && 53346.4729 &  70 &  1200 & $-$5.04&$\pm$&0.16 & OHP\\
  14155 && 53345.4242 &  84 &  1200 & $-$14.62&$\pm$&8.06 & OHP\\
        && 53345.4397 &  70 &  1200 & $-$7.55&$\pm$&6.44 & OHP\\
  17217 && 53344.4147 &  88 &   900 & $-$9.84&$\pm$&2.08 & OHP\\
        && 53344.4266 &  78 &   900 & $-$8.42&$\pm$&7.31 & OHP\\
  19257 && 53345.4570 &  81 &  1200 & 2.76&$\pm$&3.74 & OHP\\
        && 53345.4725 &  90 &  1200 & $-$1.83&$\pm$&3.14 & OHP\\
  20194 && 53344.4442 &  61 &  1200 & $-$6.54&$\pm$&2.91 & OHP\\
        && 53344.4597 &  61 &  1200 & $-$4.94&$\pm$&5.45 & OHP\\
  25021 && 53346.4909 &  96 &  1200 & $-$21.92&$\pm$&1.95 & OHP\\
        && 53346.5064 &  99 &  1200 & $-$24.64&$\pm$&1.33 & OHP\\
  26212 && 53343.5295 &  71 &  1200 & 13.71&$\pm$&6.80 & OHP\\
        && 53343.5450 &  84 &  1200 & 8.98&$\pm$&4.00 & OHP\\
        && 53344.4772 &  70 &  1200 & 9.92&$\pm$&7.21 & OHP\\
        && 53344.4926 &  61 &  1200 & 8.17&$\pm$&3.27 & OHP\\
  27464 && 53343.5621 &  72 &  1200 & $-$1.76&$\pm$&2.74 & OHP\\
        && 53343.5775 &  54 &  1200 & $-$1.00&$\pm$&1.74 & OHP\\
  30468 && 53343.5984 &  83 &  1200 & $-$13.75&$\pm$&2.27 & OHP\\
        && 53343.6141 &  88 &  1200 & $-$12.52&$\pm$&4.46 & OHP\\
        && 53346.5223 &  67 &   600 & $-$8.64&$\pm$&5.24 & OHP\\
  31489 && 53346.5353 &  67 &  1200 & 18.24&$\pm$&3.60 & OHP\\
        && 53346.5506 &  71 &  1200 & 13.00&$\pm$&2.55 & OHP\\
  38731 && 53343.6343 &  61 &  1500 & 7.76&$\pm$&2.31 & OHP\\
        && 53343.6531 &  57 &  1500 & 7.29&$\pm$&1.41 & OHP\\
  42173 && 53344.6047 &  55 &  1200 & 3.69&$\pm$&4.00 & OHP\\
        && 53344.6201 &  66 &  1200 & $-$0.83&$\pm$&4.00 & OHP\\
        && 53345.6505 &  50 &  1200 & $-$3.07&$\pm$&6.00 & OHP\\
  44372 && 53346.6125 &  34 &   153 & 12.42&$\pm$&2.00 & OHP\\
  64934 && 53345.6157 &  68 &  1200 & $-$17.26&$\pm$&2.00 & OHP\\
        && 53345.6312 &  75 &  1200 & $-$8.20&$\pm$&5.00 & OHP\\
        && 53716.6955 & 65  & 400  & $-$7.77&$\pm$&10.64 & TBL\\
        && 53716.7008 & 71  & 400  & $-$19.38&$\pm$&3.95 & TBL\\
        && 53716.7060 & 77  & 400  & $-$15.92&$\pm$&9.65 & TBL\\
        && 53716.7112 & 70  & 400  & $-$24.16&$\pm$&14.18 & TBL\\
  68725 && 53344.6354 &  51 &  1200 & $-$3.49 &$\pm$&0.52  & OHP\\
        && 53344.6509 &  79 &  1200 & $-$5.32 &$\pm$& 0.94 & OHP\\
        && 53344.6663 &  84 &  1200 & $-$8.21 &$\pm$& 1.23 & OHP\\
        && 53344.6818 &  86 &  1200 & $-$9.01 &$\pm$& 0.81 & OHP\\
        && 53345.5106 &  63 &   900 & $-$3.93&$\pm$& 2.01 & OHP\\
        && 53345.5225 &  75 &   900 & $-$3.84&$\pm$& 1.51 & OHP\\
        && 53345.5345 &  73 &   900 & $-$3.62&$\pm$& 2.05 & OHP\\
        && 53345.5465 &  62 &   900 & $-$4.19&$\pm$& 1.05 & OHP\\
        && 53345.5584 &  61 &   900 & $-$13.84&$\pm$& 4.05 & OHP\\
        && 53345.5710 &  79 &   900 & $-$13.60&$\pm$& 4.15 & OHP\\
        &&54078.6139 &127&600& $-13.36$&$\pm$&1.66&NAO \\
        &&54078.6234 &156&900& $-13.96$&$\pm$&1.67&NAO \\
        &&54078.6345 &210&600& $-07.85$&$\pm$&1.71&NAO \\
        &&54078.6422 &195&600& $-07.63$&$\pm$&1.59&NAO \\
        &&54079.6026 &113&600& $-15.93$&$\pm$&1.42&NAO \\
        &&54079.6116 &131&900& $-17.41$&$\pm$&1.52&NAO \\
        &&54079.6238 &212&900& $-10.65$&$\pm$&1.97&NAO \\
        &&54079.6336 &172&600& $-10.77$&$\pm$&1.88&NAO \\
        &&54079.6438 &200&900& $-12.19$&$\pm$&2.44&NAO \\
  81995 && 53344.7008 &  87 &  1200 & 23.00&$\pm$&1.54 & OHP\\
        && 53345.6757 &  70 &  1200 & 23.44&$\pm$&0.15 & OHP\\
        && 53345.6911 &  65 &  1200 & 25.21&$\pm$&0.66 & OHP\\
        && 53345.7065 &  72 &  1200 & 28.13&$\pm$&1.89 & OHP\\
        && 53716.6701 & 53  & 400  & $-$17.93&$\pm$&2.02 & TBL\\
        && 53716.6754 & 47  & 400  & $-$18.95&$\pm$&2.11 & TBL\\
        && 53716.6806 & 59  & 400  & $-$20.56&$\pm$&0.29 & TBL\\
        && 53716.6858 & 51  & 400  & $-$20.40&$\pm$&0.91 & TBL\\
        && 53720.6881 & 52  & 400  & 30.93&$\pm$&3.53 & TBL\\
        && 53720.6934 & 41  & 400  & 31.18&$\pm$&1.48 & TBL\\
        && 53720.6985 & 41  & 400  & 31.83&$\pm$&1.00 & TBL\\
        && 53720.7038 & 53  & 400  & 33.61&$\pm$&1.89 & TBL\\
 217860 && 53343.2347 & 108 &  1200 & 2.06&$\pm$&0.93 & OHP\\
        && 53343.2512 &  96 &  1200 & 6.14&$\pm$&1.82 & OHP\\
        && 53343.2695 & 105 &  1200 & 3.89&$\pm$&1.35 & OHP\\
 221774&A & 53345.2348 &  98 &  1200 & 21.51&$\pm$&0.38 & OHP\\
       &B&                      &       &           & $-$52.74   &$\pm$& 0.55   & \\
       &A& 53345.2502 & 113 &  1200 & 20.33&$\pm$&0.37 & OHP\\
       &    B&                      &       &           & $-$56.22   &$\pm$& 0.24  \\
       &A& 53345.2657 & 105 &  1200 & 20.90&$\pm$&0.40 & OHP\\
       &    B&                      &       &           & $-$52.80   &$\pm$& 1.80  \\
       &A& 53345.3485 &  63 &   600 & 20.71&$\pm$&0.53 & OHP\\
       &B&                      &       &           & $-$54.68   &$\pm$& 0.41  \\
       &A& 53346.2635 &  83 &   600 & 13.57&$\pm$&1.00 & OHP\\
       &B&                      &       &           & $-$58.55   &$\pm$& 1.66  \\
       &A & 53572.5836   &  95     &   500        &  $-$46.01   &$\pm$& 0.43  & TBL \\
       &B&                      &       &           & 52.72   &$\pm$& 1.96  \\
       &A & 53572.5900   &  71     &  500         &  $-$46.75   &$\pm$& 0.43   & TBL\\
       &B&                      &       &           & 50.37   &$\pm$& 1.96  \\
       &A & 53572.5963   &  83     &    500       &  $-$46.14   &$\pm$& 0.43   & TBL\\
       &B&                      &       &           & 51.14   &$\pm$& 1.96  \\
       &A & 53572.6027   &  85     &    500       &  $-$45.87   &$\pm$& 0.43   & TBL\\
       &B&                      &       &           & 49.71   &$\pm$& 1.96  \\
 223425 && 53345.2846 & 114 &  1200 & 4.61   &$\pm$& 2.5 & OHP\\
        && 53345.3001 & 112 &  1200 & 0.37   &$\pm$& 1.31 & OHP\\
 223672 && 53344.2582 &  95 &  1200 & 8.74   &$\pm$& 1.80 & OHP\\
        && 53344.2736 & 102 &  1200 & 9.09   &$\pm$& 1.80 & OHP\\
        && 53345.3184 & 118 &  1200 & 7.79   &$\pm$& 2.07 & OHP\\
        && 53345.3343 & 108 &  1200 & 5.37   &$\pm$& 3.22 & OHP\\
        && 53572.6142 & 78   & 500   & 7.66   &$\pm$& 4.50  & TBL\\
        && 53572.6206 & 77   & 500   & 3.73   &$\pm$& 3.28  & TBL\\
        && 53572.6269 & 83   &  500  & 5.47   &$\pm$& 4.13  & TBL\\
        && 53572.6333 & 74   &  500  & 2.79   &$\pm$& 2.40  & TBL\\
 224624 && 53345.3627 &  96 &  1200 & $-$9.87   &$\pm$& 1.15 & OHP\\
        && 53345.3781 &  83 &  1200 & $-$9.148   &$\pm$& 1.20 & OHP\\
        && 53346.3556 &  62 &   600 & $-$8.993   &$\pm$& 1.00 & OHP\\
        && 53573.4507 &  64 & 500   & $-$6.49   &$\pm$& 3.52 & TBL\\
        && 53573.4571 & 68  &  500  & 4.65   &$\pm$& 7.15 & TBL\\
        && 53573.4635 & 71  &  500  & $-$0.81   &$\pm$& 4.85 & TBL\\
        && 53573.4699 & 65  &  500  & $-$6.91   &$\pm$& 3.35 & TBL\\
 225125 && 53344.2914 &  89 &  1200& $-$3.25   &$\pm$& 1.60 & OHP\\
        && 53344.3068 &  93 &  1200 & $-$5.83   &$\pm$& 1.24 & OHP\\
\end{supertabular}
\end{center}

\end{document}